\documentclass[aps,prb,twocolumn,amsmath,amssymb]{revtex4}
\usepackage{graphicx} 
\usepackage{bm}
\usepackage{dcolumn}
\usepackage{amsmath}
\usepackage{epstopdf}
\usepackage{natbib}

\begin{document}

\title{Probing phase coherence via density of states for strongly correlated excitons}

\author{V. Apinyan}
\author{T. K. Kope\'{c}}
\email{v.apinyan@int.pan.wroc.pl}

\affiliation{%
Institute of Low Temperature and Structure Research (ILTSR), Polish Academy of Sciences, PO. Box 1410, 50-950 Wroclaw 2, Poland 
}%
\date{\today}

\begin{abstract}
%
We present the calculation of the coherent spectral functions and density of states (DOS) for excitonic systems in the frame of the three dimensional extended Falicov-Kimball model. By using gauge-invariant U(1) transformation to the usual fermions, we represent the electron operator as a fermion attached to the U(1) phase-flux tube. The emergent bosonic gauge field, related to the phase variables is crucial for the Bose-Einstein condensation (BEC) of excitons. Employing the path-integral formalism, we manipulate the bosonic and fermionic degrees of freedom to obtain the effective actions related to fermionic and bosonic sectors. Considering the normal and anomalous excitonic Green functions, we calculate the spectral functions, which have the forms of convolutions in the reciprocal space between bosonic and fermionic counterparts. For the fermionic incoherent part of the DOS we have found the strong evidence of the hybridization-gap in DOS spectra. Furthermore, considering Bogoliubov coherence mechanism, we calculate the coherent DOS spectra. For the coherent normal fermionic DOS, there is no hybridization-gap found in the system due to strong coherence effets and phase stiffness. The similar behavior is observed also for the condensate part of the anomalous excitonic DOS spectra. We show that for small values of the Coulomb interaction, fermionic DOS exhibits a Bardeen-Cooper-Schriffer (BCS) -like double-peak structure. In the BEC region of the BCS-BEC crossover, the double-peak structure disappears totally for both: coherent and incoherent DOS spectra. We discuss also, temperature dependence of DOS functions.
\end{abstract}
\maketitle

\section{\label{sec:Section_1} Introduction}
%
The attractive Coulomb interaction between conduction band electrons and valence band holes plays a crucial role in forming the bosonic electron-hole (e-h) bound pairs \cite{cite-1} (excitons). The Bose-Einstein condensation of the excitons represents a very interesting subject in solid state physics. Mostly, the e-h systems are realized in photoexcited semiconducting materials and the properties of such systems strongly depend on the e-h density, temperature of the system, pressure, and other important physical quantities. There is a number of works, both experimental and theoretical, where all those effects are investigated intensively. \cite{cite-1, cite-2, cite-3,cite-4, cite-5, cite-6}   
Low density system of excitons behaves like the usual Bose gas, and, at cryogenic temperatures, the BEC transition can be expected. \cite{cite-4, cite-5} On the other hand, the high density system of bound e-h pairs behaves like the system of usual Cooper pairs in superconductors. \cite{cite-7} In this limit we have the Bardeen-Cooper-Schrieffer (BCS) state of e-h pairs. \cite{cite-8, cite-9} Despite many experimental investigations to observe the coherent exciton condensates \cite{cite-10, cite-11, cite-12, cite-13, cite-14} there is not yet definitive evidence for such states. 
Therefore, an expected BCS-BEC crossover represents actually a fascinating problem typical to the excitonic systems.\cite{cite-15,cite-16,cite-17,cite-18,cite-19,cite-20} Especially, it is interesting from the viewpoint of the difference from the similar crossover in superconductors or trapped atomic Fermi gases. \cite{cite-21, cite-22, cite-23} The transition to the e-h pair condensed phase, in the limit of weak-coupling, is related to the relative motion between electrons and holes,\cite{cite-15} implying the BCS regime and is in contrast to the case of strong-coupling regime, when the BEC state is related to the motion of the center of mass of excitons. \cite{cite-15} In the whole BCS-BEC transition region, the e-h mass difference leads to a large suppression of the BEC transition temperature, which is proved to not be same as the temperature of excitonic pair (EP) formation, and hence the excitonic insulator (EI) phase. \cite{cite-15} Recently, it is shown theoretically that the excitonic insulator and the excitonic condensate are not exactly the
same states of the matter.\cite{cite-15,cite-24,cite-25} The author in Refs. \onlinecite{cite-24} and \onlinecite{cite-25} shows from general considerations that in the low density limit of the
excitonic pairs, the critical temperature of excitonic condensation should be much smaller than the
temperature of EP formation, in contrast to the previous treatments, \cite{cite-16,cite-17,cite-18,cite-19,cite-20} where the EI state is ad hoc associated with the BEC state of excitons as the same. Similarly, in Ref. \onlinecite{cite-15} it is shown that the EI state is an excitonium state, where incoherent e-h bound pairs are formed and, furthermore, at the lower temperatures, the BEC of excitons appears in consequence of the reconfiguration and coherent condensation of the preformed excitonic pairs. Obviously, in the low density limit, the gas of free excitons undergoes the BEC phase transition at the very low temperatures, and in general the BEC temperature transition line is not coinciding with that of EP formation. The Bose condensation of excitonic pairs is possible only when the macroscopic
phase coherence is attended by the system. \cite{cite-24} However, the experimental evidence
of the existence of two distinct transition temperatures, for the general case of a three dimensional (3D) bulk system, is yet lacking in the literature. In fact, the question, whether
a true, ``coherent'' BEC transition is present in the system of excitons, (in the case of high exciton density limit) is still ambiguous. The experimental proof of it, is a very cumbersome problem, because of dominant role of quantum fluctuations at low temperatures $k_{B}T\lesssim  E^{eh}_{Ry}$ (with $E^{eh}_{Ry}$ being the binding energy of a Mott-Wannier exciton), when very large zero-point oscillations are present.

The continuing growing interest, to the problem of the coherent excitonic condensates motivates us to calculate excitonic spectral functions and density of states, as a direct probe mechanism, to compare the results with the high-resolution  studies of angle-resolved photoemission spectroscopy (ARPES) \cite{cite-26,cite-27,cite-28,cite-29}, and with spectral weight measurements on the excitonic materials at the very low temperatures. 

The general strategy of our calculations is based on the effective actions method. First, we transform the initial total action of the system to a gauge invariant form, by applying the U(1) gauge transformation to the fermion operators. As a result, the electron appears in the theory like a composite object of that of the fermion
with the attached U(1) phase-``flux-tube''. The electron factorization in terms of two variables has an unprecedented impact on the whole theory. Then we integrate out the phase variables and we get the effective fermionic action in the theory and we derive a set of self-consistent equations for the EI state. Furthermore, we discuss shortly the results of the quantum rotor model obtained after the integration of fermionic degrees of freedom.

The path-integral formalism elaborated here permits to calculate the correlation functions in the system and, as a result, we obtain the expressions of normal (both incoherent and coherent) and anomalous excitonic spectral functions, and shapes of density of states (DOS) corresponding. A special attention is payed, when calculating the bosonic phase-stiffness DOS function, which is negative. Furthermore, it is crucial for calculation of the coherent normal and anomalous excitonic DOS functions. Namely, considering the expressions of the bosonic and fermionic DOS functions, we calculate the total, phase coherent DOS functions, as convolutions from bosonic and fermionic counterparts.  

For the incoherent partial normal fermionic DOS functions at $T=0$ we obtain a hybridization-gap in the excitation spectra as a direct consequence of the presence of the Hartree-type gap in the single-particle energy scales. For the case of coherent normal and anomalous excitonic DOS functions, this hybridization-gap is absent totally for all frequency modes and for all values of the Coulomb interaction. This is due of strong coherence effects in the strongly correlated fermion system at low temperatures and at low densities. For the anomalous excitonic DOS function, we found that the hybridization-gap is absent for the case of small and intermediate values of the Coulomb interaction parameter, but there is a finite constant small gap that is opening in the strong coupling regime, signaling the passage to the SC (BEC) side of the SC-SM phase transition (the BCS-BEC crossover).   

The paper is organized as follows: The paper is organized as follows:in Section \ref{sec:Section_2} we introduce the model Hamiltonian and we discuss the main calculation schemes. In Section \ref{sec:Section_3}, we describe the EI state, with a short discussion about important energy scales in the system. 
In Section \ref{sec:Section_4} we present the analytical calculation of spectral functions and DOS functions of the system in consideration. At the end of the Section \ref{sec:Section_4} we show the results of numerical evaluations for calculated DOS functions. 

%
\section{\label{sec:Section_2} The method}
%
\subsection{\label{sec:Section_2_1} The EFKM Hamiltonian}
%
As the model for study the excitonic condensation at low temperatures we have chosen the two-band extended Falicov-Kimball model (EFKM),\cite{cite-16,cite-30,cite-31,cite-32,cite-33} due to its large applicability for treatment of the electronic correlations.\cite{cite-38,cite-39} The Hamiltonian of the EFKM model is 
\begin{eqnarray}
&&{\cal{H}}=-t_{c}\sum_{\left\langle {\bf{r}},{\bf{r}}' \right\rangle}\left[\bar{c}({{\bf{r}}})c({{\bf{r}}}')+h.c.\right]-\bar{\mu}\sum_{{\bf{r}}}n({\bf{r}})-
\nonumber\\
&&-t_{f}\sum_{\left\langle {\bf{r}},{\bf{r}}' \right\rangle}\left[\bar{f}({{\bf{r}}})f({{\bf{r}}}')+h.c.\right]+\frac{\epsilon_{c}-\epsilon_{f}}{2}\sum_{{\bf{r}}}\tilde{n}({\bf{r}})+
\nonumber\\
&&+U\sum_{{\bf{r}}}\frac{1}{4}\left[n^{2}({\bf{r}})-\tilde{n}^{2}({\bf{r}})\right].
\label{Equation_1}
\end{eqnarray}
Here $\bar{f}({{\bf{r}}})$ ($\bar{c}({{\bf{r}}})$) is the $f$ ($c$) electron creation operator at the lattice position ${\bf{r}}$, the summation $\left\langle {\bf{r}}, {\bf{r}}' \right\rangle$ is over nearest neighbors (n.n) sites on the 3D cubic lattice. The short hand notations are introduced $n({\bf{r}})=n_{c}({\bf{r}})+n_{f}({\bf{r}})$ and $\tilde{n}({\bf{r}})=n_{c}({\bf{r}})-n_{f}({\bf{r}})$ in order to simplify the calculations. Next, $t_{c}$ is the hopping amplitude for $c$-electrons and $\epsilon_{c}$ is the corresponding on-site energy level parameter. Similarly, $t_{f}$ is the hopping amplitude for $f$-electrons and $\epsilon_{f}$ is the on-site energy level parameter for $f$-orbital. 
For $t_{c}t_{f}<0$ ($t_{c}t_{f}>0$) we have a direct (indirect) band gap semiconductor.
The on-site (local) Coulomb interaction $U$ in the last term of the Hamiltonian in Eq.(\ref{Equation_1}) plays the coupling role between two bands. As we will see later on, the strength of the local Coulomb interaction will tune the semi-metal (SM)- semiconductor (SC) transition in the system. The chemical potential $\bar{\mu}$ is $\bar{\mu}=\mu-\bar{\epsilon}$, where $\bar{\epsilon}=\left(\epsilon_{c}+\epsilon_{f}\right)/2$.
Naturally, we adjust the chemical potentials $\mu_{f}$ and $\mu_{c}$ in order to maintain the number of electrons in $f$ and $c$ orbitals separately. Then, the equilibrium value of the chemical potential $\mu\equiv \mu_{f}=\mu_{c}$ in ~Eq.(\ref{Equation_1}) will be determined from the half-filling condition, i.e., we suppose that $\left\langle n_{c}({\bf{r}})\right\rangle+\left\langle n_{f}({\bf{r}})\right\rangle=1$. We will use $t_{c}=1$ as the unit of energy and we fix the band parameter values $\epsilon_{c}=0$ and $\epsilon_{f}=-1$. The Fermi energy level is assumed to be situated at the level of the $c$ -band, thus $\epsilon_{F}=0$. For the $f$ -band hopping amplitude $t_{f}$ we consider the values $t_{f}=-0.3$ and $t_{f}=-0.1$ corresponding to the heavy hole and light hole $f$ -bands. Throughout the paper, we set $k_{B}=1$, $\hbar=1$ and lattice constant $a=1$. Also, we keep the frequency symbol $\nu$ for fermions and $\omega$ - for bosons, throughout the paper.  
In the case of degenerated $f$ and $c$ -bands, i.e., when $\epsilon_{f}=\epsilon_{c}$ and $t_{f}=t_{c}$, the EFKM model reduces to the standard Hubbard model.\cite{cite-36} The principal advantage of the EFKM, in comparison with the genuine Falicov-Kimball model (FKM), \cite{cite-37} is that it is taking into account the direct nearest-neighbors $f$ electron hoppings \cite{cite-38, cite-39} ($t_{f}$) and it can be shown \cite{cite-36} that the EI state is unstable when the pure FKM is approached.

In fact, the EFKM Hamiltonian in Eq.(\ref{Equation_1}) is equivalent to the asymmetric Hubbard model, if we associate to the orbitals $c$ and $f$ the spin variables, by replacing the fermionic Hilbert space with the pseudo-fermionic one and then linearizing the interaction term via the bosonic states (see in Ref.\ \onlinecite{cite-17}).
%
%
\subsubsection{\label{sec:Section_2_1_1} The partition function}
%
The Hamiltonian in Eq.(\ref{Equation_1}) is containing two separate quadratic terms and is suitable for decoupling by functional path integration method. \cite{cite-41, cite-42}
We employ the imaginary-time, fermionic path-integral method and we introduce the fermionic Grassmann variables\cite{cite-41} ${f}({{\bf{r}}}\tau)$ and ${c}({{\bf{r}}}\tau)$ at each site ${\bf{r}}$ and each time $\tau$ varying in the interval $0\leq \tau \leq\beta$, where $\beta=1/T$ with $T$ being the thermodynamic temperature.\cite{cite-43} The time-dependent variables ${f}({{\bf{r}}}\tau)$ and ${c}({{\bf{r}}}\tau)$ are satisfying the anti-periodic boundary conditions ${x}({{\bf{r}}}\tau)=-{x}({{\bf{r}}}\tau+\beta)$, where $x=f$ or $c$. 

The grand canonical partition function of system of fermions written as a functional integral over the Grassmann fields is
\begin{eqnarray}
Z_{\rm GC}=\int\left[{D}\bar{c}{D}c\right]\left[{D}\bar{f}{D}f\right]e^{-{\cal{S}}[\bar{c},c, \bar{f},f]},
\label{Equation_2}
\end{eqnarray} 
where the action in the exponent is given in the path-integral formulation in the form
\begin{eqnarray}
{\cal{S}}[\bar{c},c, \bar{f},f]=\sum_{x=f,c}{\cal{S}}_{B}[\bar{x},x]+\int^{\beta}_{0}d\tau {{H}}(\tau).
\label{Equation_3}
\end{eqnarray} 
Here, ${\cal{S}}_{B}[\bar{f},f]$ and ${\cal{S}}_{B}[\bar{c},c]$ are fermionic Berry terms. They are defined as follows
\begin{eqnarray}
{\cal{S}}_{B}[\bar{x},x]=\sum_{{\bf{r}}}\int^{\beta}_{0}d\tau \bar{x}({\bf{r}}\tau)\dot{x}({\bf{r}}\tau)
\label{Equation_4}
\end{eqnarray} 
and $\dot{x}({\bf{r}}\tau)=\partial_{\tau}x({\bf{r}}\tau)$ is the time-derivative
. 
%
\subsubsection{\label{sec:Section_2_1_2} Decoupling of term proportional to $n^{2}$}
%
We will decouple the quadratic density terms in Eq.(\ref{Equation_1}) using the Hubbard-Stratonovich linearisation procedure \cite{cite-42} and by introducing new variables $V({\bf{r}}\tau)$ and ${\cal{ \varrho}}({\bf{r}}\tau)$, conjugated respectively to the density terms $n({\bf{r}}\tau)$ and $\tilde{n}({\bf{r}}\tau)$. 
For quadratic term, proportional to $n^{2}({\bf{r}}\tau)$ in the exponent of the partition function in Eq.(\ref{Equation_2}), we have 
\begin{eqnarray}
&&\exp\left[{-\frac{U}{4}\sum_{{\bf{r}}}\int^{\beta}_{0}d\tau n^{2}\left({\bf{r}}\tau\right)}\right]
\nonumber\\
&&=\int\left[{D}V\right]e^{-\sum_{{\bf{r}}}\int^{\beta}_{0}d\tau \left[\frac{V^{2}({\bf{r}}\tau)}{U}-iV({\bf{r}}\tau)n({\bf{r}}\tau)\right]}.
\nonumber\\
\ \ \ 
\label{Equation_5}
\end{eqnarray} 
Furthermore, we combine the exponent in Eq.(\ref{Equation_5}) with the effective chemical potential term, linear in total electron density $n({\bf{r}})$ (see the second term, in Eq.(\ref{Equation_1})). Then, we decompose the variables $V({\bf{r}}\tau)$ into a static and the periodic part
\begin{eqnarray}
V({\bf{r}}\tau)=V_{0}({\bf{r}})+\tilde{V}({\bf{r}}\tau),
\label{Equation_6}
\end{eqnarray}
where, from time-periodicity of $\tilde{V}({\bf{r}}\tau)$, it follows the relation $\int^{\beta}_{0}d\tau \tilde{V}({\bf{r}}\tau)=0$. As a result, the integration over variables $V({\bf{r}}\tau)$ becomes the integration over the scalar static variables $V_{0}({\bf{r}})$ and the integration over the periodic field $\tilde{V}({\bf{r}}\tau)$:
\begin{eqnarray}
\int\left[{D}V\right]...=\int\left[{D}V_{0}\right]\int\left[{D}\tilde{V}\right]... \; .
\label{Equation_7}
\end{eqnarray}
For the periodic part in Eq.(\ref{Equation_6}), we introduce U$(1)$ phase field variables $\phi({\bf{r}}\tau)$ by using a Faraday-type relation \cite{cite-44}
\begin{eqnarray}
\tilde{V}({\bf{r}}\tau)=\frac{\partial{\phi}({{\bf{r}}}\tau)}{\partial{\tau}}\equiv \dot{\phi}({\bf{r}}\tau).
\label{Equation_8}
\end{eqnarray}
Thus, for the dynamic part, we transform the integration over the gauge variables $\tilde{{V}}({\bf{r}}\tau)$, into an integration over generic phase variables $\phi({\bf{r}}\tau)$
\begin{eqnarray}
\int\left[{\cal{D}}\tilde{V}\right]... \rightarrow \int\left[{\cal{D}}\phi\right]... .
\label{Equation_9}
\end{eqnarray}
The periodicity of $\tilde{V}\left({\bf{r}}\tau\right)$ implies that $\phi\left({\bf{r}}\beta\right)=\phi\left({\bf{r}}0\right)$.
The integration measure, in Eq.(\ref{Equation_9}), over the variables $\phi$ is defined as
\begin{eqnarray}
\int\left[{\cal{D}}\phi\right]...\equiv \int^{\infty}_{-\infty}\prod_{{\bf{r}}}d\phi_{0}({\bf{r}})
\nonumber\\
\times \int^{\phi_{f}=\phi({\bf{r}}\beta)}_{\phi_{i}=\phi_{0}({\bf{r}})}\prod_{{\bf{r}}}d\phi({\bf{r}}\tau)... , 
\label{Equation_10}
\end{eqnarray}
where the notations $\phi_{i}$ and $\phi_{f}$ mean the initial and final paths.
The path-integral in Eq.(\ref{Equation_10}) could be transformed into path integration over the compact U(1) group-manifold, since the electromagnetic group U(1), governing the phase field,
is compact, i.e., $\phi({\bf{r}}\tau)$ has the topology of a circle ($S_1$), thus we have a non-homotopic mapping of the configuration space onto the U(1) gauge group $S_{1}\rightarrow U(1)$. The paths, which
loop around a circle in different number of times, are in different
homotopy classes and they cannot be continuously deformed
into one another. All these paths can be characterized by their proper winding numbers\cite{cite-45} $m\left({\bf{r}}\right)$. Any two paths which have different winding numbers, cannot be continuously  transformed one to another, and in order to include all possible phase path contributions, we have to sum
over all topologically inequivalent phase configurations described
by their winding numbers.\cite{cite-45} Accordingly, the path-integral in Eq.(\ref{Equation_10}) is transformed as
\begin{eqnarray}
\int\left[{\cal{D}}\phi\right]...=\int\left[{\cal{D}}\varphi\right]... \; ,
\label{Equation_11}
\end{eqnarray}
where the integration measure is now
\begin{eqnarray}
\int\left[{\cal{D}}\varphi\right]...\equiv \sum_{\left\{m({\bf{r}})\right\}}\int^{2\pi}_{0}\prod_{{\bf{r}}}d\varphi_{0}({\bf{r}})
\nonumber\\
\times \int^{\varphi({\bf{r}}\beta)=\varphi_{0}\left({\bf{r}}\right)+2{\pi}m({\bf{r}})}_{\varphi\left({\bf{r}}0\right)=\varphi_{0}\left({\bf{r}}\right)}\prod_{{\bf{r}}}d\varphi({\bf{r}}\tau)...\ . 
\label{Equation_12}
\end{eqnarray}
In performing the integration over the phase field one should take into account that the field configurations satisfy the boundary conditions \cite{cite-45} 
\begin{eqnarray}
\varphi({\bf{r}}\beta)-\varphi({\bf{r}}0)=2\pi{m({\bf{r}})}.
\label{Equation_13}
\end{eqnarray}
Thus, integration over all phases $\phi({\bf{r}}\tau)$ amounts the integration over the $\beta$-periodic field $\varphi({\bf{r}}\tau)$ and the summation over a set of U(1) integer winding numbers $m({\bf{r}})$.
For the scalar static part $V_{0}({\bf{r}})$, we get the following functional integral
\begin{eqnarray}
&&\int\left[{D}V_{0}\right]e^{\sum_{{\bf{r}}}\int^{\beta}_{0}d\tau -\frac{V^{2}_{0}({\bf{r}})}{U}+iV_{0}({\bf{r}})\left[n({\bf{r}}\tau)-\frac{2\bar{\mu}}{U}\right]} \ .
\label{Equation_14}
\end{eqnarray}
The saddle-point value of ${V}_{0}({{\bf{r}}})$ is given by
\begin{eqnarray}
{{V}_{0}}=i\frac{Un}{2}-i\bar{\mu},
\label{Equation_15}
\end{eqnarray}
where $n$ is total average particle density $n=n_{c}+n_{f}$ (furthermore, we will fix $n$ as equal to $1$, corresponding to the case of half-filling). 

Thereby, after decoupling of the quadratic term proportional to $n^{2}$ in the Hamiltonian in Eq.(\ref{Equation_1}), we get a contribution to the partition function in Eq.(\ref{Equation_2}) in the form
\begin{eqnarray}
\exp\left[{-{\cal{S}}\left[\varphi\right]-\sum_{{\bf{r}}}\int^{\beta}_{0}d\tau{\mu}_{n}}n({\bf{r}}\tau)\right],
\label{Equation_16}
\end{eqnarray}
where the emergent phase-only action ${\cal{S}}[\varphi]$ is given as 
\begin{eqnarray}
{\cal{S}}[\varphi]=\sum_{{\bf{r}}}\int^{\beta}_{0}d\tau\left[\frac{\dot{\varphi}^{2}({\bf{r}}\tau)}{U}-\frac{2\bar{\mu}}{iU}\dot{\varphi}({\bf{r}}\tau)-i\dot{\varphi}({\bf{r}}\tau)n({\bf{r}}\tau)\right]
\nonumber\\
\ \ \ 
\label{Equation_17}
\end{eqnarray}
and the effective chemical potential ${\mu}_{n}$, attached to the total density operator in Eq.(\ref{Equation_16}), is given in the form ${\mu}_{n}=\frac{Un}{2}-\bar{\mu}$.
%
\subsubsection{\label{sec:Section_2_1_3} Decoupling of term proportional to $\tilde{n}^{2}$}
%
The decoupling of the quadratic term proportional to $\tilde{n}^{2}({\bf{r}}\tau)$ in the exponent of the partition function in Eq.(\ref{Equation_2}) is also straightforward. We obtain
\begin{eqnarray}
&&\exp\left[{\sum_{{\bf{r}}}\int^{\beta}_{0}d\tau \frac{U}{4}\tilde{n}^{2}({\bf{r}}\tau)}\right]
\nonumber\\
&&=\int\left[{D}{\varrho}\right]e^{-\sum_{{\bf{r}}}\int^{\beta}_{0}d\tau \left[\frac{\varrho^{2}({\bf{r}}\tau)}{U}-\varrho({\bf{r}}\tau)\tilde{n}({\bf{r}}\tau)\right]} \ .
\nonumber\\
\ \ \ 
\label{Equation_18}
\end{eqnarray}
After combining the expression in the exponent in Eq.(\ref{Equation_18}) with the similar linear term in the expression of the Hamiltonian in Eq.(\ref{Equation_1}) (see the forth-term in Eq.(\ref{Equation_1})), we have
\begin{eqnarray}
&&\int\left[{D}{\varrho}\right]e^{\sum_{{\bf{r}}}\int^{\beta}_{0}d\tau-\frac{\varrho^{2}({\bf{r}}\tau)}{4U}+\varrho({\bf{r}}\tau)\left[\tilde{n}({\bf{r}}\tau)-\frac{\epsilon_{c}-\epsilon_{f}}{2U}\right]} \ .
\nonumber\\
\label{Equation_19}
\end{eqnarray}
The saddle-point evaluation for $\varrho$ gives
\begin{eqnarray}
\varrho_{0}=\frac{U\tilde{n}}{2}-\frac{\epsilon_{c}-\epsilon_{f}}{2},
\label{Equation_20}
\end{eqnarray}
where $\tilde{n}=\left\langle \tilde{n}({\bf{r}}\tau)\right\rangle$ is the average of the particle density difference function.
As a result of the decoupling, we obtain a ``Zeeman''- like contribution to the partition function
\begin{eqnarray}
\exp\left[{-\sum_{{\bf{r}}}\int^{\beta}_{0}d\tau\mu_{\tilde{n}}\tilde{n}({\bf{r}}\tau)}\right]
\label{Equation_21}
\end{eqnarray}
with the attached effective chemical potential $\mu_{\tilde{n}}=\frac{\epsilon_{c}-\epsilon_{f}}{2}-\frac{U\tilde{n}}{2}$.
%
\subsubsection{\label{sec:Section_2_1_4} Linearized action with phase-field contribution}
%
To summarize, the grand canonical partition function of the system, after of both procedures of decoupling, is 
\begin{eqnarray}
{Z}_{GC}=\int \left[{D}\bar{c}{D}c\right]\left[{D}\bar{f}{D}f\right]\left[{D}\varphi\right]e^{-{\cal{S}}[\bar{c},c,{\bar{f}},f,\varphi]},
\label{Equation_22}
\end{eqnarray}
where the action ${\cal{S}}[\bar{c},c,{\bar{f}},f,\varphi]$ in the exponent is given by
\begin{eqnarray}
&&{\cal{S}}[\bar{c},c,{\bar{f}},f,\varphi]={\cal{S}}\left[\varphi\right]+\sum_{x=f,c}{\cal{S}}_{B}\left[\bar{x},x\right]
\nonumber\\
&&-t_{c}\sum_{\left\langle{\bf{r}}, {\bf{r}}' \right\rangle}\int^{\beta}_{0}d\tau \left[\bar{c}({{\bf{r}}}\tau)c({{\bf{r}}}'\tau)+h.c.\right]
\nonumber\\
&&-{t}_{f}\sum_{\left\langle {\bf{r}}, {\bf{r}}' \right\rangle}\int^{\beta}_{0}d\tau \left[\bar{f}({{\bf{r}}}\tau)f({{\bf{r}}}'\tau)+h.c.\right]
\nonumber\\
&&\ +\sum_{{\bf{r}}}\int^{\beta}_{0}d\tau \left[{\mu}_{n}n({\bf{r}}\tau)+\mu_{\tilde{n}}\tilde{n}({\bf{r}}\tau)\right].
\label{Equation_23}
\nonumber\\
\end{eqnarray}
After the Hubbard-Stratanovich linearisation, we got the total action of the system that is linear in terms of fermion densities and contains in addition a phase-dependent term ${\cal{S}}\left[\varphi\right]$, and also the terms, proportional to the effective chemical potentials $\mu_{n}$ and $\mu_{\tilde{n}}$.
%
\subsection{\label{sec:Section_2_2} The U$(1)$ transformation}
%
In the perspective to treat the local and non-local correlations in our excitonic system, it is important to separate the U(1) gauge degrees of freedom related to the phase sector. To this end, we perform the local gauge transformation to new fermion Grassmann variables $\tilde{f}({\bf{r}}\tau)$ and $\tilde{c}({\bf{r}}\tau)$. Meanwhile, this procedure will automatically eliminates also the last imaginary term, appearing in the expression of the phase action in Eq.(\ref{Equation_17}). 

\begin{figure}
\begin{center}
\includegraphics[width=200px,height=190px]{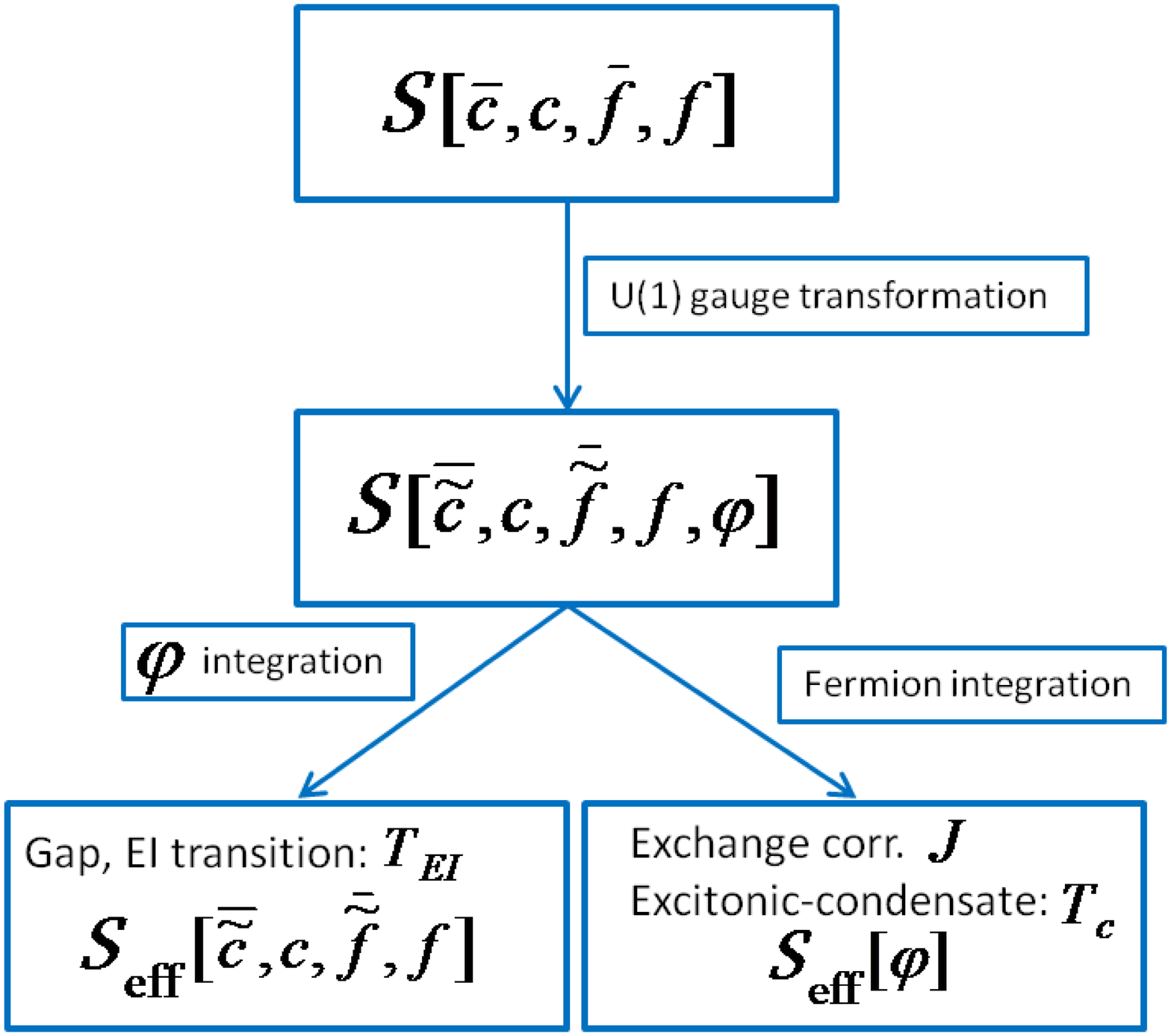}
\caption{\label{fig:Fig_1}(Color online) Functional integration procedure.}
\end{center}
\end{figure}

For the electrons of $f$ and $c$-orbitals, the U$(1)$ transformation is 
\begin{eqnarray}\left[
\begin{array}{cc}
x({\bf{r}}\tau) \\
\bar{x}({\bf{r}}\tau)
\end{array}
\right]=\hat{{\cal{U}}}(\varphi)\cdot\left[
\begin{array}{cc}
\tilde{x}({\bf{r}}\tau) \\
\bar{\tilde{x}}({\bf{r}}\tau)
\end{array}
\right],
\label{Equation_24}
\end{eqnarray}
where $\hat{\cal{U}}(\varphi)$ is the U(1) transformation matrix $\hat{\cal{U}}(\varphi)=\hat{I}\cdot\cos\varphi({\bf{r}}\tau)+i\hat{\sigma}_{z}\cdot\sin\varphi({\bf{r}}\tau)$ with the unit matrix $\hat{I}$ and $\hat{\sigma}_{z}$ being the Pauli matrix. The variables $\tilde{x}=\tilde{f}$, $\tilde{c}$, and we used the bosonic phase variables $\varphi$ introduced in Eqs.(\ref{Equation_11}) and (\ref{Equation_12}). 
After transformations given in Eq.(\ref{Equation_24}), we obtain the total action of the system in the U(1) gauge invariant form (for comparison, see the action in Eq.(\ref{Equation_23}) before transformations)
\begin{eqnarray}
&&{\cal{S}}[\bar{\tilde{c}},\tilde{c},{\bar{\tilde{f}}}, \tilde{f},\varphi]={\cal{S}}_{0}[\varphi]+\sum_{x=\tilde{f},\tilde{c}}{\cal{S}}_{B}\left[\bar{\tilde{x}},\tilde{x}\right]
\nonumber\\
&&-t\sum_{\left\langle{\bf{r}},{\bf{r}}' \right\rangle}\int^{\beta}_{0}d\tau \left[\bar{\tilde{c}}({{\bf{r}}}\tau)\tilde{c}({{\bf{r}}}'\tau)e^{-i\left[\varphi({{\bf{r}}}\tau)-\varphi({{\bf{r}}}'\tau)\right]}+h.c.\right]
\nonumber\\
&&-\tilde{t}\sum_{\left\langle {\bf{r}},{\bf{r}}' \right\rangle}\int^{\beta}_{0}d\tau \left[\bar{\tilde{f}}({{\bf{r}}}\tau)\tilde{f}({{\bf{r}}}'\tau)e^{-i\left[\varphi({{\bf{r}}}\tau)-\varphi({{\bf{r}}}'\tau)\right]}+h.c.\right]
\nonumber\\
&&\ \ \ +\sum_{{\bf{r}}}\int^{\beta}_{0}d\tau \left[{\mu}_{n}n({\bf{r}}\tau)+\mu_{\tilde{n}}\tilde{n}({\bf{r}}\tau)\right].
\nonumber\\
&&\ \ \ 
\label{Equation_25}
\end{eqnarray}
Now, $\tilde{t}$ and $t$ in Eq.(\ref{Equation_25}) are respectively $\tilde{f}$ -band and $\tilde{c}$ -band fermion transfer integrals. We got in Eq.(\ref{Equation_25}) also, a new, emergent, quadratic phase action ${\cal{S}}_{0}[\varphi]$
\begin{eqnarray} {\cal{S}}_{0}[\varphi]=\sum_{{\bf{r}}}\int^{\beta}_{0}d\tau\left[\frac{\dot{\varphi}^{2}({\bf{r}}\tau)}{U}-\frac{2\bar{\mu}}{iU}\dot{\varphi}({\bf{r}}\tau)\right].
\label{Equation_26}
\end{eqnarray}
The partition function of the system in the new variables $\tilde{f}$ and $\tilde{c}$ is 
\begin{eqnarray}
{Z}_{GC}=\int\left[D\bar{\tilde{c}}D\tilde{c}\right]\left[D\bar{\tilde{f}}D\tilde{f}\right]\left[D\varphi\right] e^{-{\cal{S}}[\bar{\tilde{c}},\tilde{c},{\bar{\tilde{f}}}, \tilde{f},\varphi]}\ .
\nonumber\\
\ \ \ 
\label{Equation_27}
\end{eqnarray}
From this form of the partition function we will generate the effective actions for fermions and for bosonic phase sector (see the general procedure presented in Fig.~\ref{fig:Fig_1}). 
%
%
\section{\label{sec:Section_3} Effective action for fermions}
%
By following left-lowest root, presented in Fig.~\ref{fig:Fig 1}, we will integrate out the phase variables. We obtain%
\begin{eqnarray}
{\cal{Z}}=\int \left[{\cal{D}}\bar{\tilde{c}}{\cal{D}}\tilde{c}\right]\left[{\cal{D}}\bar{\tilde{f}}{\cal{D}}\tilde{f}\right]e^{-{\cal{S}}_{\rm eff}[\bar{\tilde{c}},\tilde{c},{\bar{\tilde{f}}},f]} \ ,
\label{Equation_28}
\end{eqnarray}
where the effective, phase-averaged fermionic action in the exponent is
\begin{eqnarray}
{\cal{S}}_{\rm eff}[\bar{\tilde{c}},\tilde{c},{\bar{\tilde{f}}},\tilde{f}]=-\ln{\int\left[{\cal{D}}\varphi\right]}e^{-{\cal{S}}[\bar{\tilde{c}},\tilde{c},{\bar{\tilde{f}}},\tilde{f},\varphi]}.
\label{Equation_29}
\end{eqnarray}
The Fourier transformation of fermionic variables $\tilde{f}$ and $\tilde{c}$ is 
\begin{eqnarray}
x({\bf{r}}\tau)=\frac{1}{{\beta{N}}}\sum_{{\bf{k}},\nu_{n}}x_{{\bf{k}}}(\nu_{n})e^{i({\bf{k}}{\bf{r}}-\nu_{n}\tau)}
\label{Equation_30}
\end{eqnarray}
where $N$, is the number of lattice sites, and $\nu_{n}={\pi(2n+1)}/{\beta}$ are the Fermi-Matsubara frequencies \cite{cite-43} with $n=0,\pm1,\pm2,...$ .

Then, the effective phase-averaged fermionic action of the system in the Fourier space takes the following form
\begin{eqnarray}
&&{\cal{S}}_{\rm eff}\left[\tilde{\bar{c}},\tilde{c},\tilde{\bar{f}},\tilde{f}\right]
=\frac{1}{\beta{N}}\sum_{{\bf{k}}\nu_{n}}\left[\bar{\tilde{c}}_{\bf{k}}(\nu_{n}),\bar{\tilde{f}}_{\bf{k}}({\nu_{n}})\right]\times
\nonumber\\
&&\times{\cal{G}}^{-1}({\bf{k}},\nu_{n})\left[\begin{array}{cc}
{\tilde{c}}_{\bf{k}}(\nu_{n})\\
{\tilde{f}}_{\bf{k}}(\nu_{n})
\end{array}
\right].
\label{Equation_31}
\end{eqnarray}
Here, ${\cal{G}}^{-1}({\bf{k}},\nu_{n})$ is the inverse of the Green function matrix
\begin{eqnarray}
{\cal{G}}^{-1}({\bf{k}},\nu_{n})=
\left(
\begin{array}{cc}
{E}^{\tilde{c}}_{{\bf{k}}}(\nu_{n})
 & -\bar{\Delta}  \\
-\Delta & {E}^{\tilde{f}}_{{\bf{k}}}(\nu_{n})
\end{array}
\right),
\nonumber\\
\ \ \ \ 
\label{Equation_32}
\end{eqnarray}
where we have introduced the local excitonic order parameter $\Delta=\left\langle \bar{\tilde{c}}({\bf{r}}\tau)\tilde{f}({\bf{r}}\tau)\right\rangle$. The single-particle Bogoliubov quasienergies ${E}^{\tilde{f}}_{{\bf{k}}}(\nu_{n})$ and ${E}^{\tilde{c}}_{{\bf{k}}}(\nu_{n})$ are ${E}^{\tilde{f}}_{{\bf{k}}}(\nu_{n})=\bar{\epsilon}_{\tilde{f}}-i\nu_{n}-\tilde{t}_{{\bf{k}}}$ and ${E}^{\tilde{c}}_{{\bf{k}}}(\nu_{n})=\bar{\epsilon}_{\tilde{c}}-i\nu_{n}-{t}_{{\bf{k}}}$. Next, $\tilde{t}_{{\bf{k}}}$ and ${t}_{{\bf{k}}}$ are band-renormalized hopping amplitudes $\tilde{t}_{{\bf{k}}}=2\tilde{t}{\mathrm{g}}_{B}\gamma_{{\bf{k}}}$ and ${t}_{{\bf{k}}}=2t{\mathrm{g}}_{B}\gamma_{{\bf{k}}}$, where ${\mathrm{g}}_{B}$ is the 3D bandwidth renormalization factor 
\begin{eqnarray}
\mathrm{g}_{B}=\left.\left\langle e^{-i[\varphi({{\bf{r}}}\tau)-\varphi({{\bf{r}}}'\tau)]} \right\rangle\right|_{|{\bf{r}}-{\bf{r}}'|={{d}}} \ ,
\label{Equation_33}
\end{eqnarray}
and $\gamma_{{\bf{k}}}$ is the 3D lattice dispersion $\gamma_{{\bf{k}}}=\cos(k_{x}d_{x})+\cos(k_{y}d_{y})+\cos(k_{z}d_{z})$, with $d_{\alpha}$ ($\alpha=x,y,z$), being the components of the lattice spacing vector ${\bf{d}}={\bf{r}}-{\bf{r}}'$ with ${\bf{r}}$ and ${\bf{r}}'$, being nearest neighbors site positions. For the simple cubic lattice we have $d_{\alpha}\equiv a=1$.  

The quasiparticle energies $\bar{\epsilon}_{\tilde{f}}$ and $\bar{\epsilon}_{\tilde{c}}$ are of Hartree-type and they are defined in the theory by relation $\bar{\epsilon}_{\tilde{x}}=\epsilon_{{x}}-\mu+Un_{\tilde{y}}+i\left\langle\dot{\varphi}({{\bf{r}}}\tau)\right\rangle$, where $\tilde{y}$ means orbital, opposite to $\tilde{x}$.
The EI low-temperature phase is characterized by the local excitonic order parameter $\Delta$. Without any loss of generality, we can suppose the case of the EI state, with uniform real gap parameter $\Delta=\bar{\Delta}$. The EI state develops from local on-site electron-hole correlations. The expectation value, given in the expression of local EI order parameter, could be calculated in the frame of path-integral method, as well as the fermion density averages of the respective band levels $n_{\tilde{x}}=\left\langle\bar{\tilde{x}}({\bf{r}}\tau)\tilde{x}({\bf{r}}\tau)\right\rangle$. We get a set of the coupled self-consistent equations for the EI order parameter $\Delta$, single-particle fermion densities $n_{\tilde{x}}$, and EI chemical potential $\mu$
\begin{eqnarray}
&&\frac{1}{N}\sum_{{\bf{k}}}\left[f(E^{+}_{{\bf{k}}})+f(E^{-}_{{\bf{k}}})\right]=1,
\label{Equation_34} 
\newline\\
&&\tilde{n}=\frac{1}{N}\sum_{{\bf{k}}}\xi_{{\bf{k}}}\cdot\frac{f(E^{+}_{{\bf{k}}})-f(E^{-}_{{\bf{k}}})}{\sqrt{\xi^{2}_{{\bf{k}}}+4\Delta^{2}}},
\label{Equation_35} 
\newline\\
&&\Delta=-\frac{U\Delta}{N}\sum_{{\bf{k}}}\frac{f(E^{+}_{{\bf{k}}})-f(E^{-}_{{\bf{k}}})}{\sqrt{\xi^{2}_{{\bf{k}}}+4\Delta^{2}}}.
\label{Equation_36}  
\end{eqnarray}
Here $f(\epsilon)=1/\left(e^{\beta\epsilon}+1\right)$ is the Fermi - Dirac distribution function, $\xi_{{\bf{k}}}=-{t}_{{\bf{k}}}+\bar{\epsilon}_{\tilde{c}}+\tilde{t}_{{\bf{k}}}-\bar{\epsilon}_{\tilde{f}}$ is the quasiparticle dispersion and the energy parameters $E^{+}_{{\bf{k}}}$ and $E^{-}_{{\bf{k}}}$ are defined as
\begin{eqnarray}
E^{\pm}_{{\bf{k}}}=\frac{1}{2}\left(-{t}_{{\bf{k}}}+\bar{\epsilon}_{\tilde{c}}-\tilde{t}_{{\bf{k}}}+\bar{\epsilon}_{\tilde{f}}\pm{\sqrt{\xi^{2}_{{\bf{k}}}+4\Delta^{2}}}\right).
\label{Equation_37}
\end{eqnarray}
In fact, the difference between Eqs.(\ref{Equation_34}) - (\ref{Equation_36}) and the Hartree-Fock results,\cite{cite-20} is in the presence of bandwidth renormalization factor $g_{B}$. The calculation of the factor $g_{B}\left({\bf{r}}-{\bf{r}}'\right)$ could be done effectively within the self-consistent-harmonic-approximation (SCHA) method.\cite{cite-47,cite-48, cite-49} In this approximation the quantum rotor description is reduced to classical Hamiltonian one by the Feynmann-Kleinert minimization procedure. \cite{cite-47,cite-48} Our results for SCHA show that the factor $g_{B}$ is equal identically to $1$ at $T=0$ as in the two-dimensional (2D) case (see in Ref. \onlinecite{cite-47} for details). For higher temperatures, it differs from unity, but not much.

The difference between the energy parameters in Eq.(\ref{Equation_37}) defines the charge transfer gap in the system $\Delta_{c}=E^{+}_{{\bf{k}}}-E^{-}_{{\bf{k}}}=\sqrt{\xi^{2}_{{\bf{k}}}+4\Delta^{2}}$ (see in Ref. \onlinecite{cite-51} for details).

The numerical solution of the system of equation Eqs.(\ref{Equation_34}) - (\ref{Equation_36}) is discussed in details in Ref.(\onlinecite{cite-51}), where finite-difference approximation method is used in numerical evaluations. Different values of $\tilde{t}$ hopping amplitude are considered there. The results are coinciding well with the previous Hartree - Fock (HF), improved slave-boson, and 2D constrained path Monte Carlo investigations. \cite{cite-15,cite-16,cite-17,cite-18,cite-19, cite-20, cite-31,cite-32} This good correspondence is ascribed to a rather weak band renormalization at $T=0$ (in our case $g_{B}=1$). At finite temperatures, the particle number fluctuations are important and the band renormalization becomes necessary, especially, when approaching from the band insulator (BI) high-temperature side.  
Indeed, as the numerical evaluations show, the transition temperature $T_{EI}$ of the e-h pair formation is not vanishing for the case of the vanishing narrow band hopping $\tilde{t}=0$. 

The exact solutions for the chemical potential could be obtained from Eqs.(\ref{Equation_34}) - (\ref{Equation_36}), both at the boundary of EI transition (i.e., when $\Delta\left(T_{EI}, U\right)=0$) and in the EI state region ($\Delta\left(T < T_{EI}, U\right)\neq0$). The results are discussed in Ref.(\onlinecite{cite-51}).
\begin{figure}
\begin{center}
\includegraphics[scale=.6]{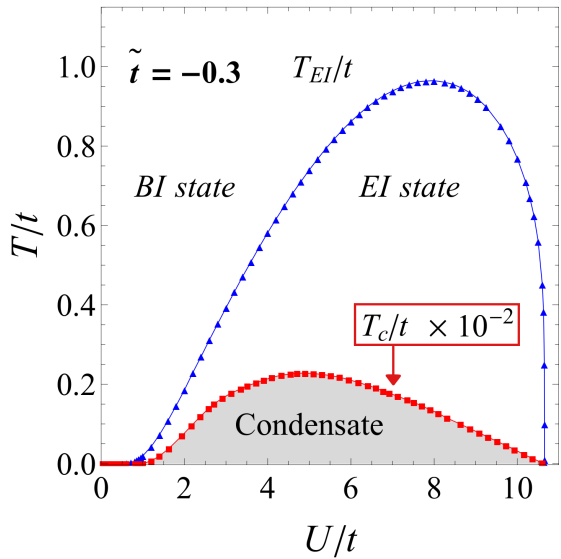}
\caption{\label{fig:Fig_2}(Color online) EI transition critical temperature $T_{EI}/t$, (upper curve) and excitonic BEC transition critical temperature $T_{c}$ (lower curve). The data for the lower curve was taken from Ref.\onlinecite{cite-51}.}
\end{center}
\end{figure}
 
Meanwhile, it is also shown in Ref.(\onlinecite{cite-51}) that the excitonic BEC transition critical temperature $T_{c}$ is much smaller than the critical temperature $T_{EI}$ of the excitonic pair formation, in good agreement with previous theoretical predictions.\cite{cite-15, cite-23, cite-24, cite-25, cite-47, cite-52} The self-consistent numerical solutions for $T_{EI}$ and $T_{c}$ are shown in Figs.~\ref{fig:Fig_2} for $\tilde{t}=-0.3$. It is clear that, for the case of intermediate and strong interaction limits, the coherent BEC transition critical temperature $T_{c}$ is smaller of about two orders of magnitude than temperature $T_{EI}$. Contrary, in the very small interaction case, we have the coincidence of both transition temperatures. This result is expected also from general considerations, because at small $U/t$ we are in the BCS limit, which means that the pairing and condensation occur simultaneously. 

A similar effect is found recently in Ref.\onlinecite{cite-47}, considering 2D excitonic systems, where the exciton-superfluid transition critical temperature is found much smaller than the critical temperature $T_{EI}$ of exciton pair formation. Such a reduction of the transition temperature, due to the coherent pairing scheme, is given also in Ref.\onlinecite{cite-52}, where the BCS-Bose crossover is studied in the 2D attractive Hubbard model. Especially, a HF pair formation temperature is estimated, which corresponds to the regime of the incoherent or local pairs (for a comparison see the correspondence to our $T_{EI}$, and order parameter $\Delta$, which is also local), in difference with superconducting pairing temperature, at which the coherent Cooper pairs start to be formed.  
The general idea used in Ref. (\onlinecite{cite-51}) is based on the non local n.n. excitonic exchange correlation mechanism and Bogoliubov's mean-field self-consistency assumption \cite{cite-53} for the effective phase action obtained after fermion integration procedure in the action in Eq.(\ref{Equation_27}) (this is represented by the right-lowest root in Fig.~\ref{fig:Fig_1}, given in the Section \ref{sec:Section_2_2}). The obtained Hamiltonian for the phase sector is very similar to the classical Hamiltonian one, \cite{cite-47,cite-48,cite-49,cite-51} with an effective phase-stiffness parameter $J$ that emerges after the fermion Wick averaging procedure.\cite{cite-43} The corresponding phase action is \cite{cite-51} 
\begin{eqnarray}
{\cal{S}}_{J}\left[\varphi\right]=-2J{\mathrm{g}_{B}}\cdot\int^{\beta}_{0}d\tau \sum_{\left\langle{\bf{r}},{\bf{r}}'\right\rangle}\cos{\left[\varphi({\bf{r}}\tau)-\varphi({\bf{r}}'\tau)\right]}.
\label{Equation_38}
\end{eqnarray}
It appears that the non-zero value of quantity $J$ is directly related with the pairing gap $\Delta$, since it is shown \cite{cite-51} that $J$ vanishes, when $\Delta=0$. Here, we present only the final analytical result for $J$ (for details see in Ref.(\onlinecite{cite-51}))
\begin{eqnarray}
&&J=\frac{16\Delta^{2}t\tilde{t}}{z^{2}{N^{2}}}\sum_{{\bf{k}},{\bf{k}}'}\frac{\epsilon\left({{\bf{k}}}\right)\epsilon\left({{\bf{k}}}'\right)}{{\sqrt{\xi^{2}_{{\bf{k}}}+4\Delta^{2}}}}\left[\Lambda_{1}({\bf{k}},{\bf{k}}')\tanh\left(\frac{\beta E^{+}_{{\bf{k}}}}{2}\right)\right.
\nonumber\\
&&\left.-\Lambda_{2}({\bf{k}},{\bf{k}}')\tanh\left(\frac{\beta E^{-}_{{\bf{k}}}}{2}\right)\right],
\nonumber\\
\label{Equation_39}
\end{eqnarray}
where $z$ is the number of n.n. sites on the 3D lattice. The parameters $\Lambda_{1}({\bf{k}},{\bf{k}}')$ and $\Lambda_{2}({\bf{k}},{\bf{k}}')$ entering in Eq.(\ref{Equation_38}) are given by $\Lambda_{1}({\bf{k}},{\bf{k}}')=\left(E^{+}({\bf{k}}) - E^{+}({\bf{k}}')\right)^{-1}\cdot\left({E^{+}({\bf{k}}) - E^{-}({\bf{k}}')}\right)^{-1}$, $\Lambda_{2}({\bf{k}},{\bf{k}}')=\left({E^{-}({\bf{k}}) - E^{-}({\bf{k}}')}\right)^{-1}\cdot\left({E^{-}({\bf{k}}) - E^{+}({\bf{k}}')}\right)^{-1}$.
Also, $J$ is strictly positive for all the regions of the normalized Coulomb interaction parameter $U/t$. In addition, it follows from the analytical form of $J$, that the macroscopic phase coherence in the system is characterized by an energy scale $J\sim t_{e}\cdot t_{h}/(t_{e}+t_{h})$ for all values of the Coulomb interaction parameter $U$. This is related to the motion of the center of mass of e-h composed quasiparticle,\cite{cite-15} because $t_{e}\cdot t_{h}/(t_{e}+t_{h}) \sim (m_{e}+m_{h})^{-1}$. For the strong interaction case, we are converging with the hard core Boson model, with the kinetic energy scale $\Delta t_{e}\cdot t_{h}/U$ (with $\Delta$ being the local excitonic order parameter). Thereby, it is shown in Ref. (\onlinecite{cite-51}) that non-local correlations between the electrons and holes of different n.n. excitonic pairs, are related with the excitonic BEC condensation. Furthermore, in the frame of the quantum rotor model, the excitonic BEC transition probability function is derived and its temperature dependence is found.\cite{cite-51} 

As a systematic continuation of theoretical study given in Ref. (\onlinecite{cite-51}), we elaborate here on the analytical forms of normal, $f$ and $c$ -band, Green functions (incoherent and coherent) and coherent, anomalous excitonic Green function. Also, we will discuss in details, the spectral functions and density of states (DOS) corresponding, in the next. The numerical evaluations of calculated DOS functions are shown in the Section \ref{sec:Section_4_5}. 
%
\section{\label{sec:Section_4} Single-particle DOS}
%
\subsection{\label{sec:Section_4_1} Normal and excitonic DOS functions}
%
The spectral density functions of the system of interacting exciton gas will determine the excitonic center-of-mass distribution related to the condensation in the low-temperature limit. Therefore the calculation of these functions represents an important task. Within our theoretical approach we can access a variety of correlation functions in the system. We will concentrate now on the $c$ and $f$ -band normal spectral functions and also the excitonic anomalous spectral function, represented in terms of the initial operators. The $c$ and $f$ -band normal excitonic Green functions are 
\begin{eqnarray}
G_{\rm x,x}({\bf{r}}\tau,{\bf{r}}'\tau')=-\langle {x}({\bf{r}}\tau)\bar{x}({\bf{r}}'\tau')\rangle,
\nonumber\\
\label{Equation_40}
\end{eqnarray}
the anomalous excitonic Green function is defined as 
\begin{eqnarray}
G_{\rm c,f}({\bf{r}}\tau,{\bf{r}}'\tau')=\langle \bar{c}({\bf{r}}\tau)f({\bf{r}}'\tau')\rangle.
\label{Equation_41}
\end{eqnarray}
After introducing the U(1) transformations, defined in Eq.(\ref{Equation_24}), we will have the Green function's decomposition as 
\begin{eqnarray}
G_{\rm x,x}({\bf{r}}\tau,{\bf{r}}'\tau')=-\langle \tilde{x}({\bf{r}}\tau)\bar{\tilde{x}}({\bf{r}}'\tau')\rangle\cdot\langle e^{-i\left[\varphi({\bf{r}}\tau)-\varphi({\bf{r}}'\tau')\right]}\rangle,
\nonumber\\
\label{Equation_42}
\newline\\
G_{\rm c,f}({\bf{r}}\tau,{\bf{r}}'\tau')=\langle \bar{\tilde{c}}({\bf{r}}\tau)\tilde{f}({\bf{r}}'\tau')\rangle\cdot\langle e^{-i\left[\varphi({\bf{r}}\tau)-\varphi({\bf{r}}'\tau')\right]}\rangle.
\nonumber\\
\label{Equation_43}
\end{eqnarray}
The phase factors, appearing in the definitions of Green functions, define in fact a charge-bosonic propagator $G_{z}({\bf{r}}\tau, {\bf{r}}'\tau')$ as follows
\begin{eqnarray}
G_{z}({\bf{r}}\tau, {\bf{r}}'\tau')=\langle e^{-i\left[\varphi({\bf{r}}\tau)-\varphi({\bf{r}}'\tau')\right]}\rangle.
\label{Equation_44}
\end{eqnarray}
Then, we pass to the Fourier space representation for the U(1) - transformed Green functions
\begin{eqnarray}
&&\tilde{G}_{\rm \tilde{x},\tilde{x}}({\bf{r}}\tau, {\bf{r}}'\tau')=\frac{1}{\beta{N}}\sum_{{\bf{k}},\nu_{n}}\tilde{G}_{\rm \tilde{x},\tilde{x}}({\bf{k}},\nu_{n})\times 
\nonumber\\
&&\times e^{i\left[{\bf{k}}({\bf{r}}-{\bf{r}}')-\nu_{n}(\tau-\tau')\right]},
\nonumber\\
\label{Equation_45}
\newline\\
&&\tilde{G}_{\rm \tilde{c},\tilde{f}}({\bf{r}}\tau, {\bf{r}}'\tau')=\frac{1}{\beta{N}}\sum_{{\bf{k}},\nu_{n}}\tilde{G}_{\rm \tilde{c},\tilde{f}}({\bf{k}},\nu_{n})\times
\nonumber\\
&&\times e^{i\left[{\bf{k}}({\bf{r}}-{\bf{r}}')-\nu_{n}(\tau-\tau')\right]}
\nonumber\\
\label{Equation_46}
\end{eqnarray}
and 
\begin{eqnarray}
G_{z}({\bf{r}}\tau, {\bf{r}}'\tau')=\frac{1}{\beta{N}}\sum_{{\bf{k}},\omega_{n}}G_{z}({\bf{k}},\omega_{n})e^{i\left[{\bf{k}}({\bf{r}}-{\bf{r}}')-\omega_{n}(\tau-\tau')\right]}.
\nonumber\\
\label{Equation_47}
\end{eqnarray}
Furthermore, the Fourier transformation of the functions in Eqs.(\ref{Equation_42})-(\ref{Equation_43}) will be rewritten as a convolution in the reciprocal ${\bf{k}}$-space
\begin{eqnarray}
G_{\rm x,x}({\bf{k}},\nu_{n})=\frac{1}{\beta{N}}\sum_{{\bf{q}},\omega_{n}}G_{z}({\bf{q}},\omega_{n})\cdot\tilde{G}_{\rm \tilde{x},\tilde{x}}({\bf{k}}-{\bf{q}},\nu_{n}-\omega_{n})
\nonumber\\
\label{Equation_48}
\end{eqnarray}
and 
\begin{eqnarray}
G_{\rm c,f}({\bf{k}},\nu_{n})=\frac{1}{\beta{N}}\sum_{{\bf{q}},\omega_{n}}G_{z}({\bf{q}},\omega_{n})\cdot\tilde{G}_{\rm \tilde{c},\tilde{f}}({\bf{k}}-{\bf{q}},\nu_{n}-\omega_{n}).
\nonumber\\
\label{Equation_49}
\end{eqnarray}
It is worth to mention that frequency summations in Eqs.(\ref{Equation_48}) and (\ref{Equation_49}) are over Bose-Matsubara frequencies $\omega_{n}={2\pi{n}/\beta}$.  
The Fermionic Green functions will be calculated using the formalism discussed in the Sections \ref{sec:Section_2} and \ref{sec:Section_3} and also, functional derivation techniques.\cite{cite-41} Particularly, for the $f$ and $c$ -band Green functions $\tilde{G}_{\rm \tilde{x},\tilde{x}}({\bf{k}},\nu_{n})$, we get 
\begin{eqnarray}
\tilde{G}_{\rm \tilde{x},\tilde{x}}\left({\bf{k}},i\nu_{n}\right)=\langle \tilde{ x}({\bf{k}},\nu_{n})\bar{\tilde{x}}({\bf{k}},\nu_{n})\rangle=
\nonumber\\
=\frac{E^{\tilde{y}}_{{\bf{k}}}\left(\nu_{n}\right)}{E^{\tilde{x}}_{{\bf{k}}}\left(\nu_{n}\right)E^{\tilde{y}}_{{\bf{k}}}\left(\nu_{n}\right)-\Delta^{2}}.
\label{Equation_50}
\end{eqnarray}
In general, the experimental observation of hybridization between the valence band and conduction band could be done by examining the ARPES spectra, which measure the spectral intensities just above and below the temperature $T_{EI}$ of excitonic pair formation. In ARPES experiments, one observes the imaginary part of the real retarded Green function, therefore the calculation of it represents a remarkable importance.
Indeed, the single-particle density of states is related with the imaginary part of the retarded Green functions, thus we need to calculate real retarded function, which corresponds to the normal Matsubara Green function $\tilde{G}_{\rm \tilde{x},\tilde{x}}\left({\bf{k}},i\nu_{n}\right)$. This could be done by the analytical continuation into the upper-half complex semi-plane ($\nu_{n}>0$) of frequency modes $i \nu_{n}$
\begin{eqnarray}
\tilde{G}^{R}_{\rm \tilde{x},\tilde{x}}({\bf{k}},\nu)=\tilde{G}_{\rm \tilde{x},\tilde{ x}}\left({\bf{k}},i\nu_{n}\right)|_{i\nu_{n}\rightarrow \nu+i\eta}.
\label{Equation_51}
\end{eqnarray}
The single-particle DOS is defined then as
\begin{eqnarray}
\rho_{\rm \tilde{x},\tilde{x}}\left({\bf{k}},\nu\right)=-\frac{1}{\pi}\operatorname{Im}\tilde{G}^{R}_{\rm x,x}({\bf{k}},\nu)=\left(\bar{\epsilon}_{\tilde{y}}-\tilde{t}_{{\bf{k}}}-\nu\right)^{2}\times
\nonumber\\
\times\delta\left[\left(\nu^{2}+A_{\bf{k}}\nu+B_{{\bf{k}}}\right)\cdot\left(\bar{\epsilon}_{\tilde{y}}-\tilde{t}_{{\bf{k}}}-\nu\right)\right],
\label{Equation_52}
\end{eqnarray}
where 
\begin{eqnarray}
&&A_{\bf{k}}=t_{{\bf{k}}}+\tilde{t}_{\bf{k}}-\bar{\epsilon}_{\tilde{x}}-\bar{\epsilon}_{\tilde{y}}
\label{Equation_53}
\end{eqnarray}
and 
\begin{eqnarray}
&&B_{\bf{k}}=\bar{\epsilon}_{\tilde{x}}\bar{\epsilon}_{\tilde{y}}+4t\tilde{t}\left[\cos(k_x)+\cos(k_y)+\cos(k_z)\right]^{2}-
\nonumber\\
&&-2\tilde{t}\bar{\epsilon}_{\tilde{x}}\left[\cos(k_x)+\cos(k_y)+\cos(k_z)\right]-
\nonumber\\
&&-2t\bar{\epsilon}_{\tilde{y}}\left[\cos(k_x)+\cos(k_y)+\cos(k_z)\right]-\Delta^{2}.
\label{Equation_54}
\end{eqnarray}
${\bf{k}}$-summed DOS will be 
\begin{eqnarray}
\rho_{\rm \tilde{x},\tilde{x}}\left(\nu\right)=\frac{1}{N}\sum_{\bf{k}}\rho_{\rm \tilde{x},\tilde{x}}\left({\bf{k}},\nu\right).
\label{Equation_55}
\end{eqnarray}

The summations over the wave vectors in Eq.(\ref{Equation_55}) can be simplified by introducing the
appropriate DOS function for the 3D cubic lattice $\rho_{3D}(x)=\frac{1}{N}\sum_{{\bf{k}}}\delta(x-\gamma_{\bf{k}})$. Then, it is not difficult to show that 
\begin{widetext}
\begin{eqnarray} 
\rho_{\rm \tilde{x},\tilde{x}}\left(\nu\right)=\int^{+3.0}_{-3.0}dx \rho_{ 3D}(x)\frac{\left[\bar{\epsilon}_{\tilde{y}}-\tilde{t}(x)-\nu\right]^{2}}{\sqrt{\xi^{2}(x)+4\Delta^{2}}}\cdot\left\{\frac{\delta\left[\nu-E^{+}(x)\right]}{|\bar{\epsilon}_{\tilde{y}}-\tilde{t}(x)-E^{+}(x)|}+\frac{\delta\left[\nu-E^{-}(x)\right]}{|\bar{\epsilon}_{\tilde{y}}-\tilde{t}(x)-E^{-}(x)|}\right\}.
\label{Equation_56}
\end{eqnarray}
\end{widetext}
Here, $\tilde{t}(x)=2\tilde{t}x$ and $t(x)=2tx$, and the energy parameters $E^{\pm}(x)$, are continuous versions of parameters defined in Eq.(\ref{Equation_37}). The density of state $\rho_{3D}(x)$, for the simple cubic 3D lattice, is given by
\begin{eqnarray}
\rho_{3D}(x)=\frac{1}{\pi^{3}}\int^{\min(1,2-x)}_{\max(-1,-2-x)}dy \frac{\Theta\left(1-\frac{|x|}{3}\right)}{\sqrt{1-y^{2}}}
\nonumber\\
\times{{\bf{K}}\left[\sqrt{1-\left(\frac{y}{2}+\frac{x}{2}\right)^{2}}\right]},
\label{Equation_57}
\nonumber\\
\ \ \ 
\end{eqnarray}
where $\Theta(x)$ is the Heaviside step function, and ${\bf{K}}(x)$ is the elliptic function of the first kind: ${\bf{K}}(x)=\int^{\pi/2}_{0}dt 1/\sqrt{1-x^{2}\sin^{2}t}$.

For the anomalous excitonic Green function, we have 
\begin{eqnarray}
\tilde{G}_{\tilde{c},\tilde{f}}\left({\bf{k}},i\nu_{n}\right)=\langle \bar{\tilde{c}}({\bf{k}},\nu_{n})\tilde{f}({\bf{k}},\nu_{n})\rangle=
\nonumber\\
=\frac{\Delta}{E^{\tilde{x}}_{{\bf{k}}}\left(\nu_{n}\right)E^{\tilde{y}}_{{\bf{k}}}\left(\nu_{n}\right)-\Delta^{2}}.
\label{Equation_58}
\end{eqnarray}
The retarded function, which corresponds to it, is then \cite{cite-43}
\begin{eqnarray}
\tilde{G}^{R}_{\tilde{c},\tilde{f}}({\bf{k}},\nu)=\tilde{G}_{\tilde{c},\tilde{f}}\left({\bf{k}},i\nu_{n}\right)|_{i\nu_{n}\rightarrow \nu+i\eta} \ .
\label{Equation_59}
\end{eqnarray}
The single-particle DOS is forthcoming then as
\begin{eqnarray}
\rho_{\tilde{c},\tilde{f}}\left({\bf{k}},\nu\right)=-\frac{1}{\pi}\operatorname{Im} \tilde{G}^{R}_{\tilde{c},\tilde{f}}({\bf{k}},\nu)=
\nonumber\\
=\Delta\cdot
\delta\left(\nu^{2}+A_{\bf{k}}\nu+B_{{\bf{k}}}\right).
\nonumber\\
\label{Equation_60}
\end{eqnarray}
Obviously, it has more simple form than the function in Eq.(\ref{Equation_52}). The ${\bf{k}}$-summed DOS for excitons will be
\begin{eqnarray}
\rho_{\rm \tilde{c},\tilde{f}}(\nu)=\Delta \cdot\left\{\frac{\rho_{3D}\left[\Lambda_{1}(\nu)\right]}{|\chi_{1}\left[\Lambda_{1}(\nu)\right]|}+\frac{\rho_{3D}\left[\Lambda_{2}(\nu)\right]}{|\chi_{2}\left[\Lambda_{2}(\nu)\right]|}\right\},
\label{Equation_61}
\end{eqnarray}
where the dimensionless parameters $\Lambda_{1,2}(\nu)$ are given by following expressions
\begin{widetext}
\begin{eqnarray}
\Lambda_{1}(\nu)=\frac{-\left[\left(t+\tilde{t}\right)\nu-\left(\bar{\epsilon}_{\tilde{c}}\tilde{t}+\bar{\epsilon}_{\tilde{f}}t\right)\right]+\sqrt{\left[\left(t-\tilde{t}\right)\nu+\left(\bar{\epsilon}_{\tilde{c}}\tilde{t}-\bar{\epsilon}_{\tilde{f}}t\right)\right]^{2}+4t\tilde{t}|\Delta|^{2}}}{4t\tilde{t}},
\label{Equation_62}
\newline\\
\Lambda_{2}(\nu)=\frac{-\left[\left(t+\tilde{t}\right)\nu-\left(\bar{\epsilon}_{\tilde{c}}\tilde{t}+\bar{\epsilon}_{\tilde{f}}t\right)\right]-\sqrt{\left[\left(t-\tilde{t}\right)\nu+\left(\bar{\epsilon}_{\tilde{c}}\tilde{t}-\bar{\epsilon}_{\tilde{f}}t\right)\right]^{2}+4t\tilde{t}|\Delta|^{2}}}{4t\tilde{t}}
\label{Equation_63}
\nonumber\\
\end{eqnarray}
\end{widetext}
and the functions $\chi_{i}\left[\Lambda_{1}(\nu)\right]$ ($i=1,2$), in the denominators in the right-hand side in Eq.(\ref{Equation_61}) are  

\begin{eqnarray}
\chi_{1}\left[\Lambda_{1}(\nu)\right]=2\left(t+\tilde{t}\right)\nu+8t\tilde{t}\Lambda_{1}(\nu)-2\left(\bar{\epsilon}_{\tilde{c}}\tilde{t}+\bar{\epsilon}_{\tilde{f}}t\right)
\nonumber\\
\label{Equation_64} 
\end{eqnarray}
and
\begin{eqnarray}
\chi_{2}\left[\Lambda_{2}(\nu)\right]=2\left(t+\tilde{t}\right)\nu+8t\tilde{t}\Lambda_{2}(\nu)-2\left(\bar{\epsilon}_{\tilde{c}}\tilde{t}+\bar{\epsilon}_{\tilde{f}}t\right).
\nonumber\\
\label{Equation_65}
\end{eqnarray}

Now, turning to the convolution forms for total fermionic and excitonic Green functions in Eqs.(\ref{Equation_48}) and (\ref{Equation_49}), we need an explicit expression for the phase-bosonic Green function $G_{z}\left({\bf{k}},\omega_{n}\right)$. We will calculate it in the formalism of the effective phase action given in the quantum rotor model, discussed earlier in Ref. \onlinecite{cite-51}, where we have derived the effective phase-only action $S_{\rm eff}[\varphi]$ by integrating the fermions in the partition function in Eq.(\ref{Equation_27}). In the following, we cast $S_{\rm eff}[\varphi]$ into the quantum rotor representation.\cite{cite-51} 
%
\subsection{\label{sec:Section_4_2} The phase-stiffness DOS}
%
To proceed, we replace the phase degrees of freedom with the complex, unimodular field $z({\bf{r}}\tau)=e^{i\varphi({\bf{r}}\tau)}$, which satisfies the time-periodic boundary condition
$z({\bf{r}}\beta)=z({\bf{r}}0)$. The spherical constraint, imposed on a set of the unimodular variables is ${1/N}\sum_{\bf{k}}|z({\bf{r}}\tau)|^{2}=1$. Now, we introduce these new variables $z({\bf{r}}\tau)$ into the partition function in Eq.(\ref{Equation_27}), in a way, consistent with the Faddeev-Popov ghost-field method \cite{cite-54}
\begin{eqnarray}
\int{\cal{D}}\bar{z}{\cal{D}}{z} \delta\left(\frac{}{}\sum_{{\bf{r}}}|z({\bf{r}}\tau)|^{2}-N\frac{}{}\right)\times
\nonumber\\
\times\delta\left(z-e^{i{\varphi}({\bf{r}}\tau)}\right)\delta\left(\bar{z}-e^{-i{\varphi}({\bf{r}}\tau)}\right)=1 \ .
\label{Equation_66}
\end{eqnarray}
The phase-phase propagator $G_{z}({\bf{r}}\tau,{\bf{r}}'\tau')$ will be rewritten in terms of $z({\bf{r}}\tau)$-variables as follows
\begin{eqnarray}
G_{z}({\bf{r}}\tau,{\bf{r}}'\tau')=\left\langle z({\bf{r}}\tau)\bar{z}({\bf{r}}'\tau')\right\rangle.
\nonumber\\
\label{Equation_67}
\end{eqnarray}
Furthermore, the variables $z({\bf{r}}\tau)$ play the role of the phase-flux attached to the fermions (see discussions in Section \ref{sec:Section_2}). In general case, the local expression of the phase-phase correlation function in Eq.(\ref{Equation_67}) is equal to unity, but, at very low temperatures (especially at $T=0$), this law breaks down, because we have to consider the symmetry breaking related to the bosonic sector, thus, critically, we have fluctuation form $z({\bf{r}}\tau)=\left\langle e^{i\varphi}({\bf{r}}\tau)\right\rangle+\tilde{z}({\bf{r}}\tau)$, and the unimodularity constraint for $z$-field is violated.

Indeed, in the very low temperature limit, considering the BEC of excitons, we have the spontaneous breaking of local U(1) gauge-symmetry, related to the phase field, leading to the non-vanishing expectation value of the $\left\langle e^{i\varphi}({\bf{r}}\tau)\right\rangle$. In order to demonstrate this, we separate the single particle states ${\bf{k}}=0$ via the Bogoliubov displacement operation (see for details in Refs.\onlinecite{cite-1}, \onlinecite{cite-51} and \onlinecite{cite-55}). This is so-called Bogoliubov phase coherence mechanism discussed in details in Ref. \onlinecite{cite-1}. Then, we write for the complex variables $z({\bf{k}},\omega_{n})$
\begin{eqnarray}
z({\bf{k}},\omega_{n})=\beta{N}\psi_{0}\delta_{{\bf{k}},0}\delta_{\omega_{n},0}+\tilde{z}({\bf{k}},\omega_{n})(1-\delta_{{\bf{k}},0})\times
\nonumber\\
\times (1-\delta_{\omega_{n},0}),
\label{Equation_68}
\end{eqnarray}
where $\psi_{0}$ is the BEC transition amplitude $\psi_{0}=\left\langle {z}({\bf{k}},\omega_{n})\right\rangle$. Next, $\tilde{z}({\bf{k}},\omega_{n})$ are the excitation part \cite{cite-58} (on-condensate) of effective bose-field.
The Fourier transformation of the phase-phase propagator $G_{z}({\bf{r}}\tau,{\bf{r}}'\tau')$ in Eq.(\ref{Equation_47}) is 
\begin{eqnarray}
G_{z}({\bf{k}}\omega_{n})=\frac{1}{\beta{N}}\left\langle z({\bf{k}},\omega_{n})\bar{z}({\bf{k}},\omega_{n})\right\rangle.
\nonumber\\
\label{Equation_69}
\end{eqnarray}
We consider the expectation value $\left\langle z({\bf{k}},\omega_{n})\bar{z}({\bf{k}},\omega_{n})\right\rangle$ in the local limit, i.e., when ${\bf{d}}={\bf{r}}-{\bf{r}}'=0$ and $\tau-\tau'=0$ and we should draw the condensate part, by applying the transformation in Eq.(\ref{Equation_68}). Hence, we have 
\begin{eqnarray}
&&G_{z}({\bf{k}},\omega_{n})=\frac{1}{\beta{N}}\left\langle z({\bf{k}},\omega_{n})\bar{z}({\bf{k}},\omega_{n})\right\rangle
\nonumber\\
&&=\beta{N}|\psi_{0}|^{2}\cdot\delta_{{\bf{k}},0}\delta_{\omega_{n},0}+\tilde G_{z}({\bf{k}},\omega_{n}).
\label{Equation_70}
\end{eqnarray}
Thereby, in Eq.(\ref{Equation_70}) we have defined the coherent macroscopic state for the excitonic system in the low temperature limit, and an excitonic BEC is expected in the next. 
The Fourier Green function $G_{z}({\bf{k}},\omega_{n})$, defined in Eq.(\ref{Equation_47}), could be calculated within the quantum rotor model, and we give here only the final result (for details, see in Ref. \onlinecite{cite-51}). 
\begin{eqnarray}
G_{z}({\bf{k}},\omega_{n})=\frac{1}{\gamma^{-1}(\omega_{n})-4J\epsilon({\bf{k}})-\lambda},
\label{Equation_71}
\end{eqnarray}
where $\gamma^{-1}(\omega_{n})$ is the inverse of the Fourier transformation of the two-point phase-phase correlation function.\cite{cite-51} We have 
\begin{eqnarray}
\gamma(\omega_{n})=\frac{8}{UZ_{0}}\sum^{+\infty}_{{m}=-\infty}\frac{e^{-\frac{U\beta}{4}\left({{m}}-\frac{2\bar{\mu}}{U}\right)^{2}}}{1-16\left[ \frac{i\omega_{n}}{U}-\frac{1}{2}\left({{m}}-\frac{2\bar{\mu}}{U}\right)\right]^{2}},
\label{Equation_72}
\nonumber\\
\ \ \ \
\end{eqnarray}
where $Z_{0}$ is the partition function of the non-interacting Bose sector
\begin{eqnarray}
Z_{0}=\sum^{+\infty}_{{m}=-\infty}e^{-\frac{U\beta}{4}\left({{m}}-\frac{2\bar{\mu}}{U}\right)^{2}}.
\label{Equation_73}
\end{eqnarray}
The summations, in Eqs.(\ref{Equation_72}) and (\ref{Equation_73}), run over topological winding numbers $m$ of the group U(1).
The retarded bosonic Green function \cite{cite-58} is related to the Matsubara Green function, by the analytical continuation
\begin{eqnarray}
G^{R}_{z}({\bf{k}},\omega)=G_{z}({\bf{k}},i\omega_{n})|_{i\omega_{n}\rightarrow \omega+i\eta} \ .
\label{Equation_74}
\end{eqnarray}
And the ${\bf{k}}$-summed DOS for bosons reads as
\begin{eqnarray}
\rho_{z}(\omega)=-\frac{1}{\pi}\sum_{{\bf{k}}}\operatorname{Im}G^{R}_{z}({\bf{k}},\omega).
\label{Equation_75}
\end{eqnarray}
After non difficult algebraic manipulations and replacing the summation in Eq.(\ref{Equation_74}) by integration with the help of 3D density of states $\rho_{3D}\left(x\right)=\frac{1}{N}\sum_{{\bf{k}}}\delta(x-\gamma_{{\bf{k}}})$ we get
\begin{eqnarray}
\rho_{z}(\omega)=-\int^{+\infty}_{-\infty}dx \frac{U\rho_{3D}\left(x\right)}{4}\cdot\left[\frac{\delta\left[\omega-\kappa_{1}(x)\right]}{\sqrt{\bar{\mu}^{2}+4UJ\left(3-x\right)}}+\right.
\nonumber\\
\left.+\frac{\delta\left[\omega-\kappa_{2}(x)\right]}{\sqrt{\bar{\mu}^{2}+4UJ\left(3-x\right)}}\right],
\label{Equation_76}
\nonumber\\
\end{eqnarray}
where $\kappa_{i}(x)$ $i=1,2$ are given by the following relations $\kappa_{1,2}(x)=-\bar{\mu}\pm{\sqrt{\bar{\mu}^{2}+4U{J}\left(3-x\right)}}$ and the stiffness parameter $J$ is given in Eq.(\ref{Equation_38}) in Section \ref{sec:Section_3}. As it could be expected the Bosonic DOS function Eq.(\ref{Equation_76}) is negative $\rho_{z}(\omega)< 0$. This is consistent with the general considerations of the weakly non-ideal Bose gas. \cite{cite-58}
%
\subsection{\label{sec:Section_4_3} Spectral density functions and fermionic DOS}
%
Furthermore, we separate the condensate modes $\left\{ {\bf{q}}={{0}},\omega_{n}=0 \right\}$ in Eqs.(\ref{Equation_48}) and (\ref{Equation_49}). We have
\begin{eqnarray}
&&G_{\rm x,x}({\bf{k}},\nu_{n})=|\psi_{0}|^{2} \cdot G_{\rm \tilde{x},\tilde{x}}({\bf{k}},\nu_{n})+
\nonumber\\
&&+\frac{1}{\beta{N}}\sum_{\substack{{\bf{q}}\neq 0 \\ \omega_{n}\neq 0}}\tilde G_{z}({\bf{q}},\omega_{n})\cdot G_{\rm \tilde{x},\tilde{x}}({\bf{k}}-{\bf{q}},\nu_{n}-\omega_{n})
\label{Equation_77} 
\end{eqnarray}
and 
\begin{eqnarray}
&&G_{\rm c,f}({\bf{k}},\nu_{n})=|\psi_{0}|^{2} \cdot G_{\rm \tilde{c},\tilde{f}}({\bf{k}},\nu_{n})+
\nonumber\\
&&+\frac{1}{\beta{N}}\sum_{\substack{{\bf{q}}\neq 0 \\ \omega_{n}\neq 0}}\tilde G_{z}({\bf{q}},\omega_{n})\cdot G_{\rm \tilde{c},\tilde{f}}({\bf{k}}-{\bf{q}},\nu_{n}-\omega_{n}).
\label{Equation_78}
\end{eqnarray}
As we see, the normal and excitonic propagators are composed of two parts, one, responsible for the condensate state and the other- the on-condensate excitation part (see discussion in Ref.\onlinecite{cite-58} for the case of the pure Bose-gas). Note also, that first terms in the right-hand sides in Eqs.(\ref{Equation_77}) and (\ref{Equation_78}) consist of the condensate-transition probability function $|\psi_{0}|^{2}$, multiplied with the fermionic propagators $G_{\rm \tilde{x},\tilde{x}}({\bf{k}},\nu_{n})$ and $G_{\rm \tilde{c},\tilde{f}}({\bf{k}},\nu_{n})$. 

Now, we are ready to calculate the analytical forms of the normal excitonic spectral functions $A_{\rm x,x}({\bf{k}},\nu)$ ($x=f,c$) and anomalous excitonic spectral function $A_{\rm c,f}({\bf{k}},\nu)$ and, later on, the profiles of the respective DOS, including states of the condensate. We introduce here the spectral functions $A_{\rm x,x}({\bf{k}},\nu)$ and $A_{\rm c,f}({\bf{k}},\nu)$ that carries the same physical information as the correlation functions $G_{\rm x,x}({\bf{k}},\nu_{n})$ and $G_{\rm c,f}({\bf{k}},\nu_{n})$. We have
\begin{eqnarray}
G_{\rm x,x}({\bf{k}},\nu_{n})=\int^{+\infty}_{-\infty}d\nu'\frac{A_{\rm x,x}({\bf{k}},\nu')}{i\nu_{n}-\nu'},
\label{Equation_79} 
\newline\\
G_{\rm c,f}({\bf{k}},\nu_{n})=\int^{+\infty}_{-\infty}d\nu'\frac{A_{\rm c,f}({\bf{k}},\nu')}{i\nu_{n}-\nu'} .
\label{Equation_80}
\end{eqnarray}
The integration here, is over continuous frequencies. Note, that $G_{\rm x,x}({\bf{k}},\nu_{n})$ and $G_{\rm c,f}({\bf{k}},\nu_{n})$ are total fermionic Green functions, including also the convolutions with bosonic parts. In the same way, we can introduce the spectral functions $A_{z}({\bf{k}},\nu)$, $A_{\rm \tilde{x},\tilde{x}}({\bf{k}},\nu)$ and $A_{\rm \tilde{c},\tilde{f}}({\bf{k}},\nu)$, associated with the charge and the pure fermionic parts (without bosonic sector). They correspond respectively to the correlation functions $G_{z}({\bf{k}},\omega_{n})$, $G_{\rm \tilde{x},\tilde{x}}({\bf{k}},\nu_{n})$ and $G_{\rm \tilde{c},\tilde{f}}({\bf{k}},\nu_{n})$. We have the following equations for these counterparts
\begin{eqnarray}
G_{z}({\bf{k}},\omega_{n})=\int^{+\infty}_{-\infty}d\nu'\frac{A_{z}({\bf{k}},\nu')}{i\omega_{n}-\nu'},
\label{Equation_81} 
\newline\\
G_{\rm \tilde{x},\tilde{x}}({\bf{k}},\nu_{n})=\int^{+\infty}_{-\infty}d\nu'\frac{A_{\rm \tilde{x},\tilde{x}}({\bf{k}},\nu')}{i\nu_{n}-\nu'}
\label{Equation_82} 
\end{eqnarray}
and 
\begin{eqnarray}
G_{\rm \tilde{c},\tilde{f}}({\bf{k}},\nu_{n})=\int^{+\infty}_{-\infty}d\nu'\frac{A_{\rm \tilde{c},\tilde{f}}({\bf{k}},\nu')}{i\nu_{n}-\nu'}.
\label{Equation_83} 
\end{eqnarray}
Using these definitions we get for the total spectral density functions $A_{\rm x,x}({\bf{k}},\nu)$ and $A_{\rm c,f}({\bf{k}},\nu)$ see Appendix \ref{sec:Section_A} 
\begin{eqnarray}
&&A_{\rm x,x}({\bf{k}},\nu)=|\psi_{0}|^{2}\cdot A_{\rm \tilde{x},\tilde{x}}({\bf{k}},\nu)-
\nonumber\\
&&-\frac{1}{N}\sum_{{\bf{q}}\neq 0}\int {d\nu'}A_{\rm z}({\bf{q}},\nu')A_{\rm \tilde{x},\tilde{x}}({\bf{k}}-{\bf{q}},\nu-\nu')\times
\nonumber\\
&&\times\left[n(\nu')+f(\nu-\nu')\right]
\nonumber\\
\label{Equation_84} 
\end{eqnarray}
and 
\begin{eqnarray}
&&A_{\rm c,f}({\bf{k}},\nu)=|\psi_{0}|^{2}\cdot A_{\rm \tilde{c},\tilde{f}}({\bf{k}},\nu)-
\nonumber\\
&&-\frac{1}{N}\sum_{{\bf{q}}\neq 0}\int {d\nu'}A_{\rm z}({\bf{q}},\nu')A_{\rm \tilde{c},\tilde{f}}({\bf{k}}-{\bf{q}},\nu-\nu')\times
\nonumber\\
&&\times\left[n(\nu')+f(\nu-\nu')\right],
\nonumber\\
\label{Equation_85} 
\end{eqnarray}
where $n(\epsilon)=1/\left(e^{\beta\epsilon}-1\right)$ is the Bose-Einstein distribution function. 
The proof of these relations is given in the Appendix \ref{sec:Section_A}.
From the spectral functions we can obtain the corresponding DOS, by summing over the reciprocal wave vectors ${\bf{k}}$, hence, the total DOS are $\rho_{\rm  x,x}(\nu)=\frac{1}{N}\sum_{\rm {\bf{k}}}A_{\rm x,x}({\bf{k}},\nu)$ and $\rho_{\rm  c,f}(\nu)=\frac{1}{N}\sum_{\rm {\bf{k}}}A_{\rm c,f}({\bf{k}},\nu)$.

Furthermore, using the expressions for $A_{\rm x,x}({\bf{k}},\nu)$ and $A_{\rm c,f}({\bf{k}},\nu)$ in Eqs.(\ref{Equation_84}) and (\ref{Equation_85}), we get for the total DOS functions
\begin{eqnarray}
\rho_{\rm x,x}(\nu)=|\psi_{0}|^{2}\cdot\rho_{\rm \tilde{x},\tilde{x}}(\nu)+\tilde{\rho}_{\rm \tilde{x},\tilde{x}}(\nu)
\label{Equation_86} 
\end{eqnarray}
and 
\begin{eqnarray}
\rho_{\rm c,f}(\nu)=|\psi_{0}|^{2}\cdot\rho_{\rm \tilde{c},\tilde{f}}(\nu)+\tilde{\rho}_{\rm \tilde{c},\tilde{f}}(\nu),
\label{Equation_87} 
\end{eqnarray}
where $\tilde{\rho}_{\rm \tilde{x},\tilde{x}}(\nu)$ and $\tilde{\rho}_{\rm \tilde{c},\tilde{f}}(\nu)$ are DOS, corresponding to the excitation part of the system and are given, as convolutions, in terms of continuous frequency modes (see Appendix \ref{sec:Section_A})
\begin{eqnarray}
&&\tilde{\rho}_{\rm \tilde{x},\tilde{x}}(\nu)=-\int^{+\infty}_{-\infty}d\nu' \rho_{z}(\nu')\rho_{\rm \tilde{x},\tilde{x}}(\nu-\nu')
\nonumber\\
&&\times \left[n\left(\nu'\right)+f\left(\nu-\nu'\right)\right]
\label{Equation_88} 
\end{eqnarray}
and
\begin{eqnarray}
&&\tilde{\rho}_{\rm \tilde{c},\tilde{f}}(\nu)=-\int^{+\infty}_{-\infty}d\nu' \rho_{z}(\nu')\rho_{\rm \tilde{c},\tilde{f}}(\nu-\nu')
\nonumber\\
&&\times \left[n\left(\nu'\right)+f\left(\nu-\nu'\right)\right].
\label{Equation_89} 
\end{eqnarray}
A key feature of the results in Eqs.(\ref{Equation_86}) and (\ref{Equation_87}), is that we have separated DOS contributions coming from the condensate and excitation parts.
We define also the total DOS function as
\begin{eqnarray}
\rho(\nu)=\sum_{x=f,c}\rho_{\rm x,x}(\nu).
\label{Equation_90}
\end{eqnarray} 
In the Section \ref{sec:Section_4_5}, we present the numerical evaluations of all discussed DOS functions.
%
\subsection{\label{sec:Section_4_5} Total DOS functions}
\begin{figure}
\includegraphics[width=210px, height=600px]{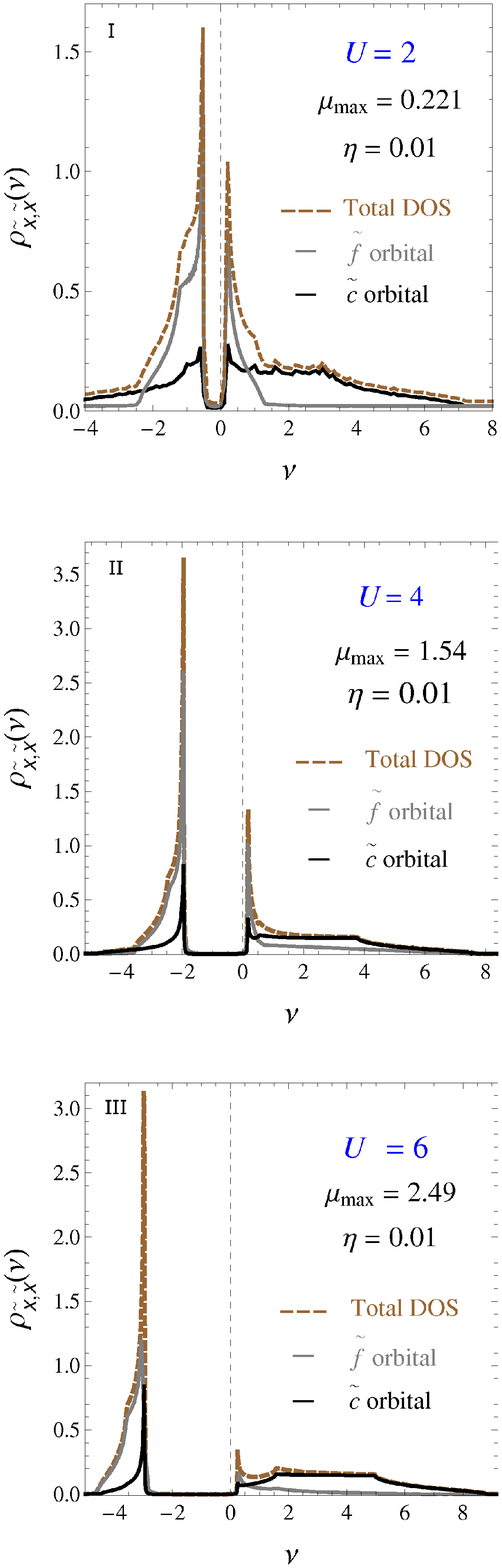}
\caption{\label{fig:Fig_3}(Color online) Single-particle normal DOS functions (incoherent) for different values of the Coulomb interaction parameter $U$ and for the case $T=0$. In panels I-III, $x=f$ and $x=c$ DOS structures are shown, and total DOS is plotted for $\tilde{t}=-0.3$.}
\end{figure}

\begin{figure}
\begin{center}
\includegraphics[width=210px, height=600px]{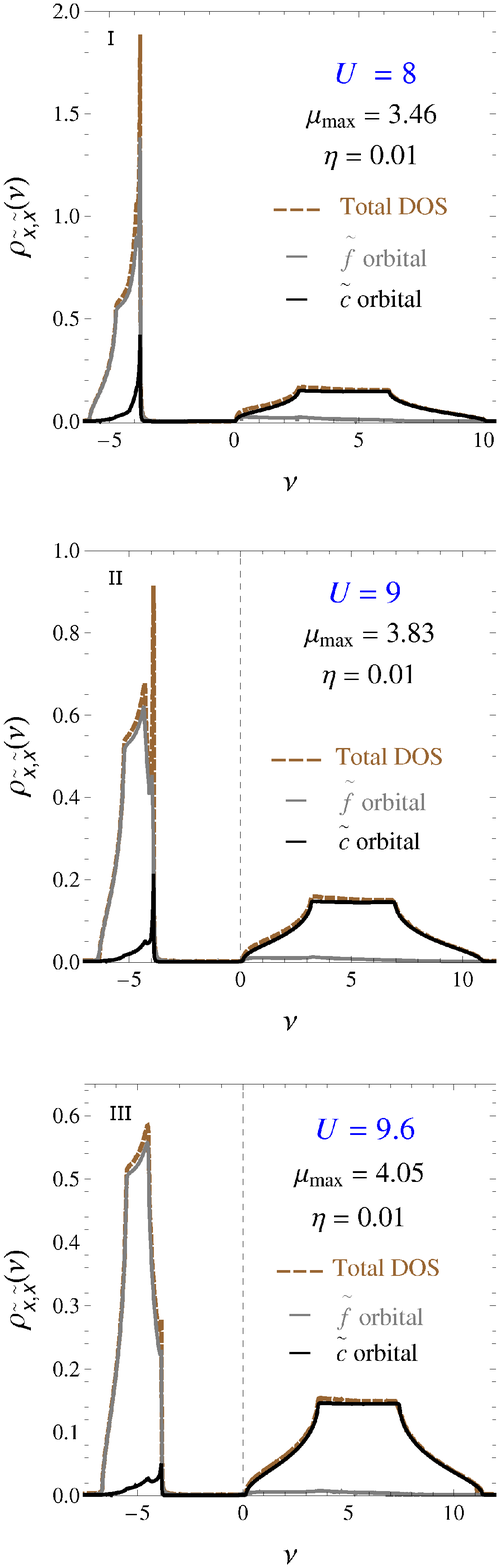}
\caption{\label{fig:Fig_4}(Color online) Single-particle normal DOS functions (incoherent) for different values of the Coulomb interaction parameter $U$ and for the case $T=0$. In panels I-III, $x=f$ and $x=c$ DOS structures are shown, and total DOS is plotted for $\tilde{t}=-0.3$.}
\end{center}
\end{figure}

\begin{figure}
\includegraphics[width=200px, height=600px]{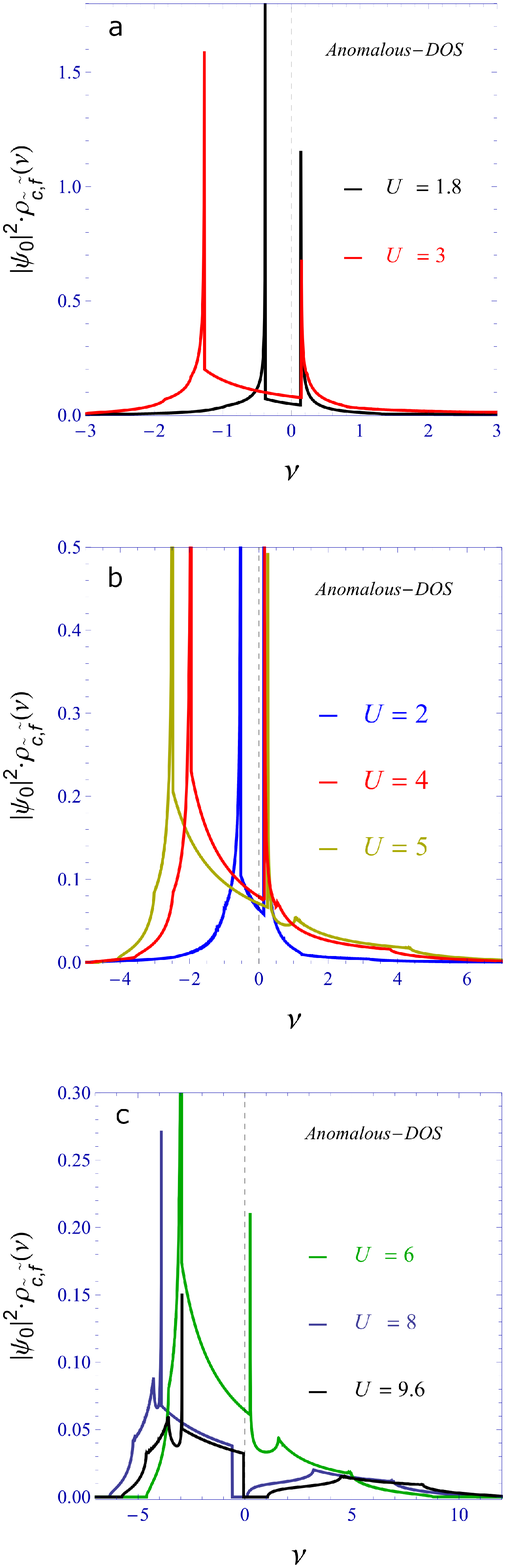}
\caption{\label{fig:Fig_5}(Color online) Condensate DOS function $|\psi_{0}|^{2}\cdot\rho_{\rm \tilde{c}\tilde{f}}$ for different values of the Coulomb interaction parameter $U$ and
for the case T = 0. The case $\tilde{t}=-0.3$ is considered here,
and the values of the functions $|\psi_{0}|^{2}$ were taken from the
work in Ref. \onlinecite{cite-51}.}
\end{figure}

\begin{figure}
\includegraphics[width=210px, height=210px]{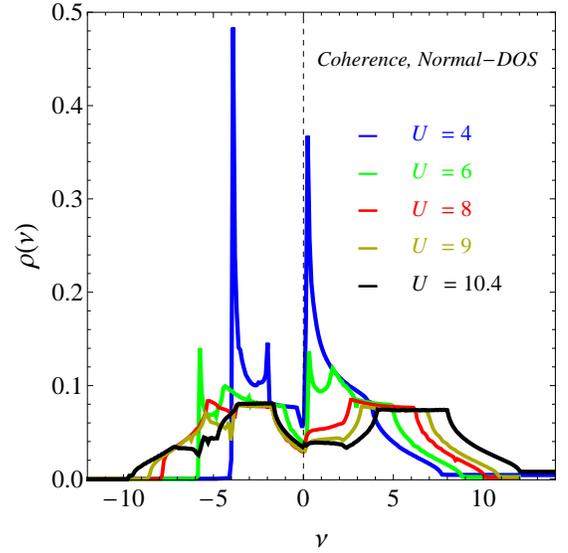}
\caption{\label{fig:Fig_6}(Color online) Total phase coherent DOS function $\rho(\nu)$ given in Eq.(\ref{Equation_90}) for different values of the Coulomb interaction parameter $U$ ($U=4$, $U=6$, $U=8$, $U=9$, $U=10.4$ in the figure) and for the case $T=0$. The case $\tilde{t}=-0.3$ is considered here, and the values of the functions $|\psi_{0}|^{2}$ were taken from the work in Ref. \onlinecite{cite-51}.}
\end{figure}

\begin{figure}
\includegraphics[width=200px, height=600px]{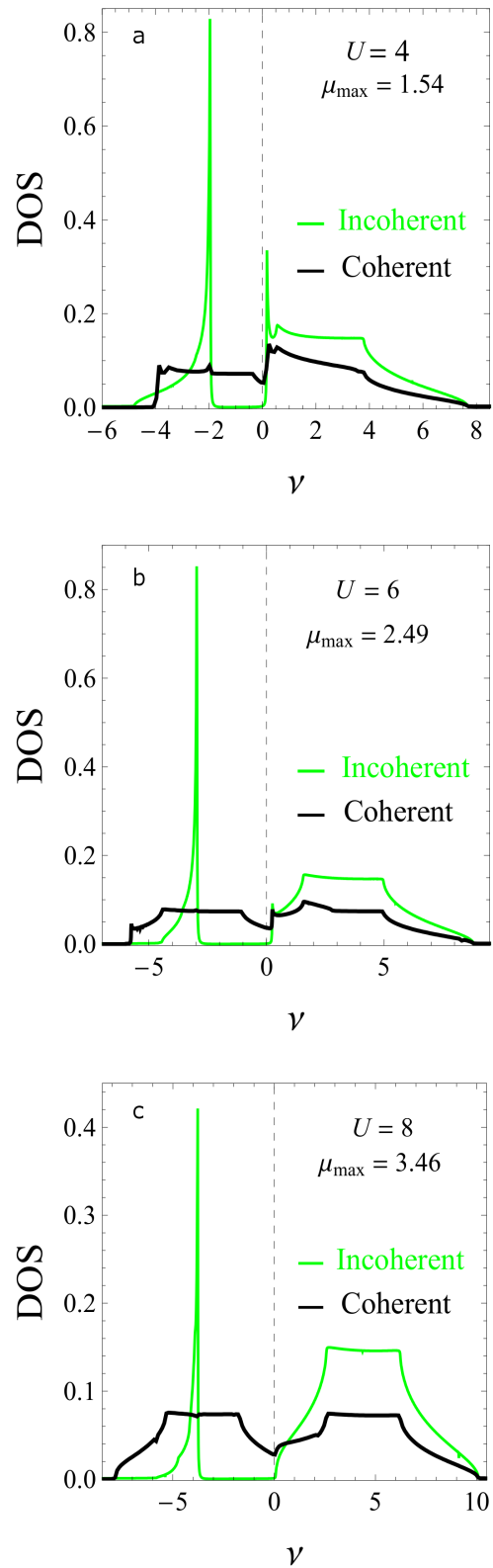}
\caption{\label{fig:Fig_7}(Color online) Incoherent and coherent normal DOS functions for the $c$ -band and for different values of the parameter $U$ ($U=4$ in a, $U=6$ in b and $U=8$ in c). The case $\tilde{t}=-0.3$ is considered here.}
\end{figure}

%
Employing Eqs.(\ref{Equation_56}),(\ref{Equation_61}) and (\ref{Equation_76}), we can obtain the explicite analytical expressions for DOS functions in Eqs.(\ref{Equation_86}) and (\ref{Equation_87}).
For the normal $f$ and $c$ -band excitonic DOS functions, we obtain
\begin{eqnarray}
&&\rho_{\rm x,x}(\nu)=|\psi_{0}|^{2}\cdot \rho_{\rm \tilde{x},\tilde{x}}(\nu)
\nonumber\\
&&-U\int^{+3}_{-3}dx \frac{\rho_{3D}(x)}{4\sqrt{\bar{\mu}^{2}+4UJ\left(3-x\right)}}\times
\nonumber\\
&&\left\{\rho_{\rm \tilde{x},\tilde{x}}\left(\nu-\kappa_{1}\left(x\right)\right)\cdot\left[n\left(\kappa_{1}(x)\right)+f\left(\nu-\kappa_{1}(x)\right)\right]+\right.
\nonumber\\
&&\left.\rho_{\rm \tilde{x},\tilde{x}}\left(\nu-\kappa_{2}\left(x\right)\right)\cdot\left[n\left(\kappa_{2}(x)\right)+f\left(\nu-\kappa_{2}(x)\right)\right]\right\}.
\label{Equation_91}
\end{eqnarray}
For the anomalous excitonic DOS function, we have
\begin{eqnarray}
&&\rho_{\rm c,f}(\nu)=|\psi_{0}|^{2}\cdot \rho_{\rm \tilde{c},\tilde{f}}(\nu)
\nonumber\\
&&-U\int^{+3}_{-3}dx \frac{\rho_{3D}(x)}{4\sqrt{\bar{\mu}^{2}+4UJ\left(3-x\right)}}\times
\nonumber\\
&&\left\{\rho_{\rm \tilde{c},\tilde{f}}\left(\nu-\kappa_{1}\left(x\right)\right)\cdot\left[n\left(\kappa_{1}(x)\right)+f\left(\nu-\kappa_{1}(x)\right)\right]+\right.
\nonumber\\
&&\left.\rho_{\rm \tilde{c},\tilde{f}}\left(\nu-\kappa_{2}\left(x\right)\right)\cdot\left[n\left(\kappa_{2}(x)\right)+f\left(\nu-\kappa_{2}(x)\right)\right]\right\}.
\label{Equation_92}
\end{eqnarray}
The numerical evaluations of calculated DOS functions at $T=0$ are given in Figs.~\ref{fig:Fig_3} - ~\ref{fig:Fig_7} for $\tilde{t}=-0.3$. The presence of singularities in the integration region causes that we used an adaptive 21-point integration routine combined with the Wynn $\epsilon$-algorithm, \cite{cite-59} to calculate those integrals numerically. The accuracy for adaptive evaluations is achieved with an absolute error of order of $10^{-4}$ and with a relative error of order of $10^{-7}$. 

Particularly, in Figs.~\ref{fig:Fig_3} and Fig.~\ref{fig:Fig_4}, we have presented purely fermionic normal single-particle (incoherent) DOS $\rho_{\rm \tilde{x},\tilde{x}}(\omega)$. We examine their behavior over the entire BCS-BEC crossover region (i.e., for different values of the Coulomb interaction $U$). An artificial Lorentzian broadening $\eta=0.01$ is used in numerical evaluations for the incoherent partial and total DOS functions of $f$ and $c$ -orbitals. The chemical potential values are inputed along the upper-bound $\mu_{max}$, where they are maximal (see in Ref.\onlinecite{cite-51}). The principal reason of it is that the BEC transition amplitude $\psi_{0}$ has no physical solutions along the lower-bound $\mu_{min}$ of the chemical potential. On the other hand, the values $\mu_{max}$ are most convenient, because they are minimalizing the Hamiltonian of the system. Furthermore, as the reference for the coherent BEC transition amplitude $|\psi_{0}|^{2}$, we considered the self-consistent calculation-results from the work in Ref.\onlinecite{cite-51}, where this function is calculated both analytically and numerically for different values of the Coulomb interaction parameter $U$ and for different temperatures, including the zero temperature limit. We use here the numerical data, which are evaluated there. 

We see in Fig.~\ref{fig:Fig_3}, that at the small interaction limit, when $2 \leqslant U\leqslant 6$ (i.e., the BCS limit), the incoherent excitonic DOS exhibits a BCS-like double-peak structure (see the panels I - III in Fig.~\ref{fig:Fig_3}) and the peaks are separated with a well defined hybridization-gap. The principal reason of it is the non-vanishing Hartree-gap $\Delta_{H}\neq 0$ in the single-particle energy-spectrum discussed above, in the Section \ref{sec:Section_3}. The hybridization-gap is proportional to the parameter $U$ and it is increasing with the increase of parameter $U$. We observe also that the peaks becomes more separated when increasing of $U$. In the strong interaction limit this displacement is stabilizing and we have practically constant value of the hybridization-gap, when further increasing the interaction (see the panels I - III, in Fig.~\ref{fig:Fig_4}).
The results in Figs.~\ref{fig:Fig_3} and ~\ref{fig:Fig_4} are very similar with the previous theoretical results. \cite{cite-15, cite-20, cite-56, cite-57} Especially, they are close to those presented in Refs.\onlinecite{cite-17} and \onlinecite{cite-20}, where the partial incoherent $f$ and $c$ -band normal DOS functions and total DOS is calculated using HF and SO(2)-invariant slave boson approaches. 
%

In Fig.~\ref{fig:Fig_5} (see the panels a - c), we have shown the coherent condensate part of the anomalous excitonic DOS given by the first term in the right-hand side in Eq.(\ref{Equation_91}), corresponding to the fundamental state (${\bf{k}}=0$). Different values of the Coulomb interaction are considered, including the small and strong interaction cases. Here, we observe again the double-peak fermionic structure, but there is no the hybridization-gap for the small and medium values of the Coulomb interaction parameter (see the panel-a and panel-b in Fig.~\ref{fig:Fig_5}), and we have finite number of states for all values of the frequency modes $\nu$. This is due to the coherence effects and the presence of the coherent excitonic condensate at the fundamental mode $\nu=0$. 
For higher values of $U$ in Fig.~\ref{fig:Fig_5} (see the panel-c), this double-peak structure in DOS is smoothing, but, in contrast to the incoherent normal DOS behavior (see in Figs.~\ref{fig:Fig_3} and ~\ref{fig:Fig_4}), here a small Mott-gap appears at the very high values of the Coulomb interaction parameter (see the DOS curves for $U=8$ and $U=9.6$ in the panel-c in Fig.~\ref{fig:Fig_5}). This is due to the fact, that the very strong Coulomb interaction has a destructive role on the condensate state and, in the large-$U$ limit of interaction, we have the destruction of the excitonic condensate, and a very small, Mott-type hybridization-gap is enhanced. In this region of the interaction, we have the coherent exciton DOS separation into two separate parts, similar to the case of the incoherent DOS functions in Figs.~\ref{fig:Fig_3} and ~\ref{fig:Fig_4}. 
%

In the positive (negative) frequency regions, the DOS spectrum in Figs.~\ref{fig:Fig_3}, ~\ref{fig:Fig_4}, and ~\ref{fig:Fig_5} is slightly displacing and broadening into the direction of higher (smaller) frequencies, for both - normal (incoherent) single-particle fermionic, and coherent excitonic DOS functions and, for both, we observe also a gradual decreases in the DOS amplitudes, across the whole BCS-BEC crossover region, when increasing the Coulomb interaction parameter $U$.   
%
%
In Fig.~\ref{fig:Fig_6}, we have shown the total coherent fermionic DOS functions $\rho(\nu)$, given by the Eq.(\ref{Equation_90}), for different values of the Coulomb interaction parameter $U$ and in the limit of zero temperature. An artificial Lorentzian broadening $\eta$ = 0.01 is again used during the numerical calculations. Contrary the case of the incoherent DOS in Figs.~\ref{fig:Fig_3} and ~\ref{fig:Fig_4}, the coherent total fermionic DOS in Fig.~\ref{fig:Fig_6} shows a different, gapless behavior. Here, again, as in the case of the incoherent DOS functions, for small and intermediate interactions $U$, we have typical double-peak fermionic structure in the DOS, which is smoothing and disappearing totally in the strong interaction limite. But, in difference with incoherent DOS functions, here we have not the presence of the hybridization-gap in the spectra and we have always have a finite number of states for all values of the frequency modes (see in Fig.~\ref{fig:Fig_6}). The reason of this gapless DOS behavior could be related to the strong coherence effects between two bands, which is due to the presence of the phase stiffness mechanism considered here. It is worth to mention that another gapless-type behavior in the DOS spectrum is found recently in Ref.\onlinecite{cite-60}, where this effect is associated with metallic charge-density-wave phase and is driven by strong electron correlations. In Fig.~\ref{fig:Fig_6}, all figures are combined together, in the way, to see the total DOS evolution with variation of the interaction parameter $U$. We see clearly, how the tuning of the interaction parameter $U$, affects the general DOS behavior in the model, by reducing the DOS amplitudes with increasing of $U$ and reducing also the number of states at the Fermi level $\rho(\nu=\epsilon_{F})$ when increasing $U$. Thereby, in the case of normal fermionic DOS (both incoherent and coherent), presented in Figs.~\ref{fig:Fig_3}, ~\ref{fig:Fig_4},~\ref{fig:Fig_5} and Fig.~\ref{fig:Fig_6}, the double-peak structure disappears for $U \gtrsim 6$, signaling the appearance of the SM (BEC) limit of the transition.
%

In Fig.~\ref{fig:Fig_7} a comparison is given for the incoherent and coherent $c$ -band normal DOS behaviors. Three different values of the Coulomb interaction $U$ are chosen (see the panels -a, b and c). We see clearly in Fig.~\ref{fig:Fig_7}, how the coherence effects are reducing the incoherent DOS amplitudes and DOS spectra become broader, along frequency axis, and also, there is no gap near the Fermi level ($\rho(\nu \sim \epsilon_{F})\neq 0$). 
 
\begin{figure}
\begin{center}
\includegraphics[scale=.35]{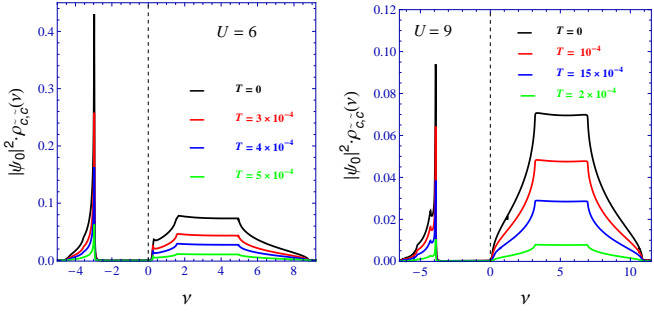}
\caption{\label{fig:Fig_8}(Color online) Temperature dependence of the normal $c$ -band DOS function for the case $\tilde{t}=-0.3$ and for different values of the Coulomb interaction parameter $U$.}
\end{center}
\end{figure}

\begin{figure}
\begin{center}
\includegraphics[scale=.3]{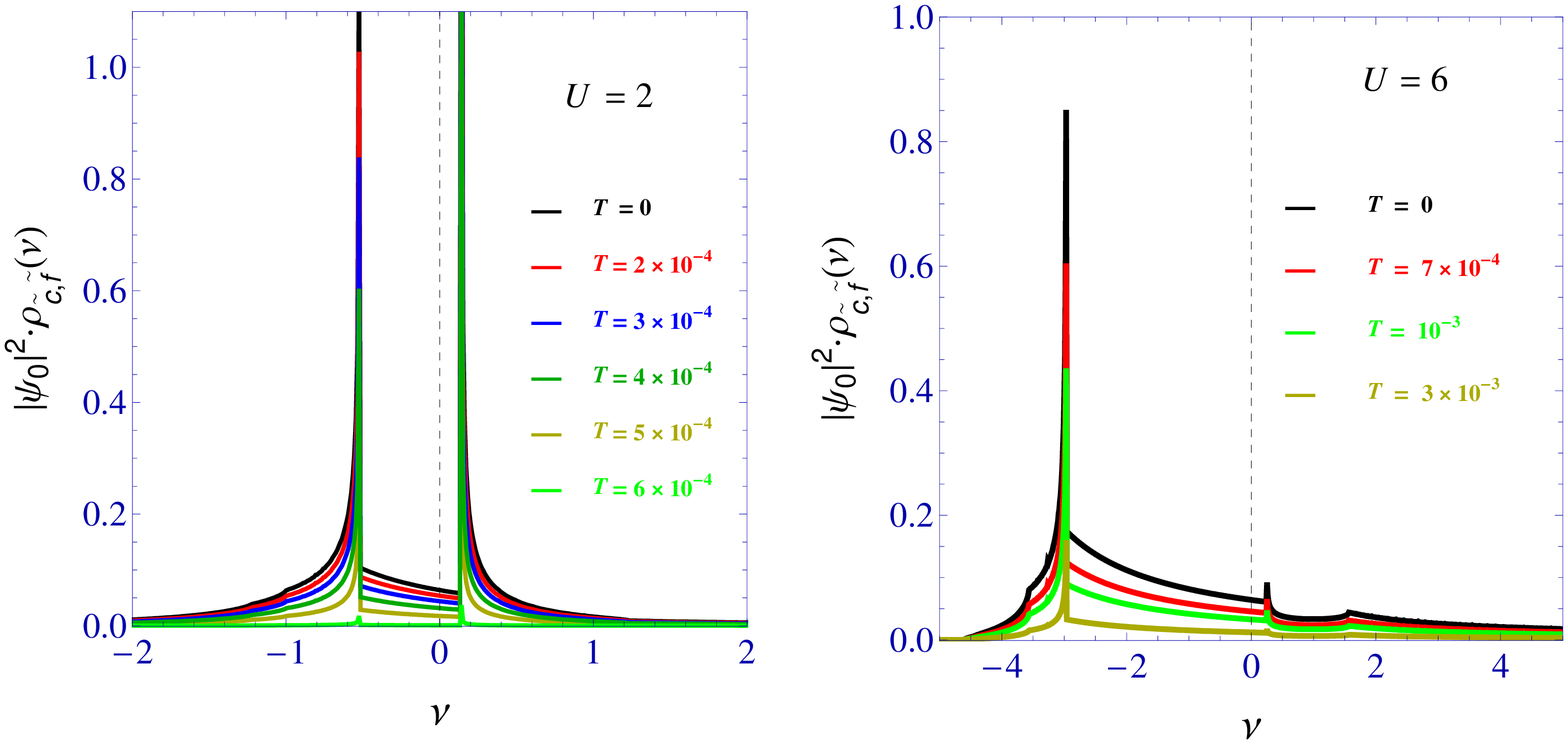}
\caption{\label{fig:Fig_9}(Color online) Temperature dependence of the anomalous excitonic DOS function for the case $\tilde{t}=-0.3$ and for different values of the Coulomb interaction parameter $U$.}
\end{center}
\end{figure}

In Figs.~\ref{fig:Fig_8} and ~\ref{fig:Fig_9}, we have presented the temperature dependence of the $c$ -band normal fermionic (Fig.~\ref{fig:Fig_8}) and anomalous excitonic DOS functions (Figs.~\ref{fig:Fig_9}). They are given by the first terms in the right-hand side in Eqs.(\ref{Equation_91}) and (\ref{Equation_92}). In the left-panel in Fig.~\ref{fig:Fig_8}, we have presented the temperature dependence of the single-particle DOS $|\psi_{0}|^{2}\rho_{\rm c,c}(\nu)$, which corresponds to the fundamental state ${\bf{k}}=0$, and for $U=6$. The case $\tilde{t}=-0.3$ is considered. The corresponding values of the amplitude $\psi_{0}$ of BEC transition, are taken again from the work in Ref.\onlinecite{cite-51}. In the right-panel in Fig.~\ref{fig:Fig_9}, the same function is plotted for the case $U=9$. As we see in Fig.~\ref{fig:Fig_8}, the hybridization-gap is still open for all values of the temperature. So the system is
always an insulator.
%

In Fig.~\ref{fig:Fig_9}, the temperature dependence of the phase-coherent anomalous excitonic condensate part of the DOS is presented (i.e., ${\bf{k}}=0$, and without excitation part) for $U=2$ and $U=6$. We observe in all Figs.~\ref{fig:Fig_8} and ~\ref{fig:Fig_9}, that the temperature has a destructive effect on the DOS amplitudes.  

We realize also, from Figs.~\ref{fig:Fig_8} and ~\ref{fig:Fig_9}, that the anomalous excitonic DOS functions vanish at the temperatures, that are far away from the region of the EI transition of about two orders of magnitude (compare the temperature scales in Figs.~\ref{fig:Fig_8} and ~\ref{fig:Fig_9} with those given in Fig.~\ref{fig:Fig_2} in the Section \ref{sec:Section_3}). This result, could be regarded as a good proof of the theory elaborated in Ref.\onlinecite{cite-51}, and we can theoretically clearly state that the excitonic BEC and EI states are not the same phases of matter. We see also, in Figs.~\ref{fig:Fig_8} and ~\ref{fig:Fig_9}, that the normal single-particle DOS persists for a rather large values of temperature, than the anomalous condensate DOS, because of the presence of the hybridization-gap in the low-energy spectra.           
%
\section{Final remarks and conclusions}

We have studied 3D system of conduction band electrons and valence band holes in the frame of the extended Falicov-Kimball model. We have implemented the path-integral formalism, in which the Coulomb interaction term is expressed in terms of U(1) quantum phase variables $\varphi$ conjugated to the local particle number, providing a useful interpretation of the problem. In Section \ref{sec:Section_3}, we have shown that at low temperatures, the electron-hole system becomes unstable with respect to the formation of the excitons at  $T=T_{EI}$, and the local gap $\Delta$ is present in the excitation spectrum, controlled by the Coulomb interaction parameter $U/t$, which gives the relevant energy scales for the excitonic insulator state. Here, as a result of the spontaneous symmetry breaking, an expectation value of $\langle e^{i\varphi}\rangle\neq 0$ appears, which is signaling of the presence of the phase coherence in the system. Furthermore, pairing and condensation are not generally the same, as it was admitted in the literature, except the weak interaction limit, when we have a BCS-like condensation of excitonic pairs.  
However, in the excitonic system with the strong pairing, we have the situation, where the pairs are strongly bound, but are uncorrelated one with each other, until they become phase coherent at temperatures $T\lesssim T_{c}$. This situation was studied in details in Ref.\onlinecite{cite-51}.

We have evaluated the normal and anomalous excitonic spectral functions for the $f$ and $c$ -band. We have determined, both analytically and numerically, density of states (DOS) spectra, governed by the pure-fermionic part (due to the condensate modes ${\bf{k}}=0$), and the excitation spectra. We have shown that there is a usual hybridization-gap in the normal (incoherent) $f$ and $c$ -band DOS structures. Contrary, in the case of the coherent normal DOS functions (presented in Fig.~\ref{fig:Fig_6}) this gap is lacking, and there is always a finite number of states at all frequency modes $\nu$. We associate this result to the strong coherence effects, that are present in the system at low temperatures.

In the anomalous excitonic DOS structure we have found that the hybridization-gap is absent for the weak and intermediate values of the Coulomb interaction parameter $U$ and this is due to the presence of the coherent excitonic condensate and strong coherence effects. A very small gap is opening in the spectra in the strong interaction limit, signaling the destruction of the coherence and condensate state.

The excitonic phase coherence may be evidenced by the coherence of their light emission, which can be studied by interferometry measurements.\cite{cite-61} Therefore, measurements  of the intensity of the line-shape of the excitons decay (by emitting photons, upon electron-hole recombination) may be a powerful probe of the DOS spectra in the excitonic systems.\cite{cite-62}

However, a final remark that should be featured is such, that for the experimental determination of the excitonic Bose-Einstein condensate, the ARPES measurements should be provided at temperatures much lower than temperatures at which the excitonic insulator state is being determined. \cite{cite-20}This is important for the achievement of macroscopic phase coherece between excitonic pairs, and the strong coherence effects are manifesting at the very low temperatures. 

As a continuation of our studies, we would like to consider the role of the charge-density-wave-like excitations in the excitonic systems, within the EFKM model and consider the temperature effects. It is especially interesting to find out how the coherent excitonic DOS will be affected in the case of presence of such elementary excitations. This will give us a more complete picture on the excitonic phase transition scenario.    

%
\section*{ACKNOWLEDGMENTS}

\appendix*
%
\section{\label{sec:Section_A_1} A convolution for DOS }
%
Here, we give a short derivation of Eqs. (\ref{Equation_84}) and (\ref{Equation_85}) presented in the Section \ref{sec:Section_4_3}. To this end, we use the denitions in Eqs.(\ref{Equation_81}) - (\ref{Equation_83}) and the convolution form of the Green functions given in Eqs.(\ref{Equation_77}) and (\ref{Equation_78}). Then, it is easy to see that the following identity holds for the integrals of the normal $x=f, c$ -band fermionic spectral functions. 
\begin{eqnarray}
&&\int d\nu \frac{A_{x,x}\left({\bf{k}},\nu\right)}{i\nu_{n}-\nu}=|\psi_{0}|^{2}\cdot\int d\nu \frac{A_{\tilde{x},\tilde{x}}\left({\bf{k}},\nu\right)}{i\nu_{n}-\nu}+
\nonumber\\
&&+\frac{1}{\beta{N}}\sum_{{\bf{q}}\neq 0, \omega_{n}\neq 0}\int d\nu\int d\nu'\left[\frac{1}{i\omega_{n}-\nu'}-\right.
\nonumber\\
&&\left.-\frac{1}{i\omega_{n}-i\nu_{n}+\nu'-\nu}\right]\cdot A_{z}\left({\bf{q},\nu'}\right)\cdot A_{\tilde{x}\tilde{x}}\left({\bf{k}}-{\bf{q}},\nu-\nu'\right).
\nonumber\\
\label{Equation_A1}
\end{eqnarray}
Furthermore, for calculating the Matsubara sums over bosonic frequencies $\omega_{n}$, we will use the property of the Bose-Einstein distribution function $n\left(\epsilon\right)$ (see the Section \ref{sec:Section_4_3}) of complex argument
\begin{equation}
n\left(i\nu_{n}+\nu-\nu'\right)=-f\left(\nu-\nu'\right),
\label{Equation_A2}
\end{equation}
where $f\left(\epsilon\right)$ is the Fermi-Dirac distribution function. In Eq.(\ref{Equation_A2}), we used the fact that $\nu_{n}$ are even Fermion-Matsubara frequencies $\nu_{n}=\frac{\pi}{\beta}\left(2n+1\right)$.  
Next, we sum the Bosonic Matsubara frequencies in Eq.(\ref{Equation_A1}) amd we rewrite the equality in Eq.(\ref{Equation_A1}) in the following form
\begin{widetext}
\begin{equation}
A_{x,x}\left({\bf{k}},\nu\right)=|\psi_{0}|^{2}\cdot A_{\tilde{x},\tilde{x}}\left({\bf{k}},\nu\right)-\frac{1}{N}\sum_{{\bf{q}}\neq 0}\int d\nu'A_{z}\left({\bf{q}},\nu'\right)A_{\tilde{x},\tilde{x}}\left({\bf{k}}-{\bf{q}},\nu-\nu'\right)\cdot \left[n\left(\nu'\right)+f\left(\nu-\nu'\right)\right].
\label{Equation_A3}
\end{equation}
\end{widetext}
The derivation of the convolution form, given in Eq.(\ref{Equation_85}), is exactly the same. 

Summing over the wave vectors ${\bf{k}}$ in Eq.(\ref{Equation_A3}), we will have the forms of the normal $f$ and $c$ -bands density of states $\rho_{x,x}\left(\nu\right)$ and also the excitonic DOS function $\rho_{c,f}\left(\nu\right)$ given in Eqs.(\ref{Equation_86}) and (\ref{Equation_87}) in the Section \ref{sec:Section_4_3}.
%
\subsection{\label{sec:Section_A_2} Normalization conditions for DOS}
%
We have a composed nature of the interacting electron, given in the Section \ref{sec:Section_2}, thus, we have various sum rules, corresponding to different counterparts of the total coherent DOS functions.
As a general rule, the partial normal DOS, and anomalous excitonic DOS functions $\rho_{x,x}\left(\nu\right)$ and $\rho_{c,f}\left(\nu\right)$ satisfy following normalization conditions
\begin{eqnarray}
\int^{+\infty}_{-\infty} d\nu \rho_{\rm f,f}\left(\nu\right)=n_{\rm f},
\label{Equation_91} 
\nonumber\\
\int^{+\infty}_{-\infty} d\nu \rho_{\rm c,c}\left(\nu\right)=n_{\rm c},
\label{Equation_A4}
\end{eqnarray}
and 
\begin{eqnarray}
\int^{+\infty}_{-\infty} d\nu \rho_{\rm c,f}\left(\nu\right)=1
\label{Equation_A5}
\end{eqnarray}
For the total normal band DOS function $\rho(\nu)$, given in Eq.(\ref{Equation_90}), we have 
\begin{eqnarray}
\int^{+\infty}_{-\infty}d\nu {\rho}\left(\nu\right)=1 \ .
\label{Equation_A6} 
\end{eqnarray}
The functions $\tilde{\rho}_{\rm \tilde{x},\tilde{x}}(\nu)$ and $\tilde{\rho}_{\rm \tilde{c},\tilde{f}}(\nu)$ in Eqs.(\ref{Equation_88}) and (\ref{Equation_89}) are excitation parts \cite{cite-58} of the respective total DOS functions in Eqs.(\ref{Equation_86}) and (\ref{Equation_87}), i.e., without condensate modes ${\bf{k}}=0$, therefore, the following normalization conditions hold  
\begin{eqnarray}
\int^{+\infty}_{-\infty} d\nu \rho_{\rm \tilde{f},\tilde{f}}(\nu)=n_{f}-|\psi_{0}|^{2},
\nonumber\\
\label{Equation_A7}
\int^{+\infty}_{-\infty} d\nu \rho_{\rm \tilde{c},\tilde{c}}(\nu)=n_{c}-|\psi_{0}|^{2}
\label{Equation_A8}
\end{eqnarray}
and 
\begin{eqnarray}
\int^{+\infty}_{-\infty} d\nu \rho_{\rm \tilde{c},\tilde{f}}(\nu)=1-|\psi_{0}|^{2} \ .
\label{Equation_A9}
\end{eqnarray}
In contrary, the integral over the charge-DOS function $\rho_{z}(\omega)$, given in Eq.(\ref{Equation_76}), vanishes, because it is antisymmetric: $\rho_{z}(\omega)=-\rho_{z}(-\omega)$. Then
\begin{eqnarray}
\int^{+\infty}_{-\infty} d\omega \rho_{z}(\omega)=0 \ .
\label{Equation_A10}
\end{eqnarray}
The norms of the full spectral density functions in Eqs.(\ref{Equation_84}) and (\ref{Equation_85}) depend on the constraints in the charge and spin bosonic sectors. 
%

\end{document}